\documentclass[11pt, superscriptaddress]{revtex4}
\usepackage{multirow}
\usepackage{graphicx}
\usepackage{stmaryrd}
\usepackage{color}

\def\rp#1{\left(#1\right)}
\def\kp#1{\left\{#1\right\}}
\def\sqp#1{\left[#1\right]}

\def\bra#1{\langle{#1}|}
\def\ket#1{|{#1}\rangle}
\def\braket#1#2{\langle{#1}|{#2}\rangle}
\def\ketbra#1#2{|{#1}\rangle\langle{#2}|}
\def\pro#1{\langle{#1}\rangle}
\def\ave#1{\langle{#1}\rangle}

\def\matrix22#1#2#3#4{\left(\begin{array}{cc}{#1} & {#2}\\{#3} & {#4}\end{array}\right)}
\def\matrixtt#1#2#3#4#5#6#7#8#9{\left(\begin{array}{ccc}{#1} & {#2} & {#3}\\{#4} & {#5} & {#6}\\{#7} & {#8} & {#9} \end{array}\right)}

\def\ba{\begin{align}}
\def\ea{\end{align}}

\begin{document}
\title{Properties of nitrogen-vacancy centers in diamond: group theoretic approach}
%\author{J. R. Maze$^{1,2,*}$ and A. Gali$^{3,*}$, E. Togan$^1$, Y. Chu$^1$, A. Trifonov$^1$, E. Kaxiras$^1$ and M.D. Lukin$^1$}

\author{J. R. Maze}
\email{jmaze@puc.cl}
\affiliation{Department of Physics, Harvard University, Cambridge, MA 02138, USA} %
\affiliation{Facultad de F\'{i}sica, Pontificia Universidad Cat\'{o}lica de Chile, Casilla 306, Santiago, Chile} %
\author{A. Gali}
\email{agali@eik.bme.hu}
\affiliation{Department of Atomic Physics, Budapest University of
  Technology and Economics, Budafoki \'ut 8, H-1111 Budapest, Hungary}%
\affiliation{Research Institute for Solid State Physics and Optics,
  Hungarian Academy of Sciences, PO Box 49, H-1525, Budapest, Hungary}%
\author{E. Togan}
\affiliation{Department of Physics, Harvard University, Cambridge, MA 02138, USA} %
\author{Y. Chu}
\affiliation{Department of Physics, Harvard University, Cambridge, MA 02138, USA} %
\author{A. Trifonov}
\affiliation{Department of Physics, Harvard University, Cambridge, MA 02138, USA} %
\author{E. Kaxiras}
\affiliation{Department of Physics and School of Engineering and Applied Sciences, Harvard University, Cambridge, MA 02138, USA} %
\author{M. D. Lukin}
\affiliation{Department of Physics, Harvard University, Cambridge, MA 02138, USA} %

\begin{abstract}
We present a procedure that makes use of group theory to analyze and
predict the main properties of the negatively charged nitrogen-vacancy
(NV) center in diamond. We focus on the relatively low temperatures
limit where both the spin-spin and spin-orbit effects are important to
consider. We demonstrate that group theory may be used to clarify
several aspects of the NV structure, such as ordering of the singlets
in the ($e^2$) electronic configuration, the spin-spin and the
spin-orbit interactions in the ($ae$) electronic configuration. We
also discuss how the optical selection rules and the response of the
center to electric field can be used for spin-photon entanglement
schemes. Our general formalism is applicable to a broad class of local
defects in solids. The present results have important implications for
applications in quantum information science and nanomagnetometry.
\end{abstract}

\maketitle

\date{}

\section{Introduction}

During the past few years nitrogen-vacancy (NV) centers have emerged
as promising candidates for a number of applications
\cite{Taylor:NatPhys2008,Degen:APL2008,Maze:Nature2008,Balasubramanian:Nature2008}
ranging from high spatial resolution imaging
\cite{Rittweger:NatPhot2009} to quantum computation
\cite{Wrachtrup:JPCM2006}. At low temperatures, the optical
transitions of the NV center become very narrow and can be coherently
manipulated, allowing for spin-photon entanglement generation
\cite{Togan:Nature2010} for quantum communication and all optical
control \cite{Santori:PRL2006}. A detailed understanding of the
properties of this defect is critical for many of these
applications. Several studies have addressed this issue both
experimentally \cite{Manson:PRB2006,Batalov:PRL2009} and theoretically
\cite{Lenef:PRB1996,Rogers:NJP2009}. Furthermore, other atom-like
defects can potentially be engineered in diamond
\cite{Aharonovich:NL2009} and other materials with similar or perhaps
better properties suitable for the desired application. Therefore, it
is of immediate importance to develop a formalism to analyze and
predict the main properties of defects in solids.

Here we present a formalism based on a group theoretical
description. While we focus on describing the nitrogen-vacancy center
in diamond, our formalism can be applied to any point defect in solid
state physics. Our method takes advantage of the symmetry of the
states to properly treat the relevant interactions and their
symmetries. We apply group theory to find out not only the symmetry of
the eigenstates but also their explicit form in terms of orbital and
spin degrees of freedom.  We show that this is essential to build an
accurate model of the NV center. In particular, we analyze the effect
of the Coulomb interaction and predict that the ordering of the
triplet and singlet states in the ground state configuration is
$\kp{{^3A_2},{^1E},{^1A_1}}$ and that the distance between them is on
the order of the exchange term of the electron-electron Coulomb
energy. This ordering has been debated over the last few years and our
results agree with recent \emph{ab initio} calculations carried in
bulk diamond \cite{Ma:PRB2010}. 

Our method is also used to analyze
important properties of the center such as polarization selection rules. The
explicit form of the states allows us to identify a particularly useful
lambda-type transition that was recently used for spin-photon entanglement generation\cite{Togan:Nature2010}. %We show that from the $A_2({^3E})$ excited state the electron can decay to the ${^3A_2}$ $m_s=1$ ($m_s=-1$) ground state by emitting a right (left) circularly polarized photon as a consequence of the spin-orbit interaction. This process is particularly robust due to the spin-spin interaction in the excited state. 
We also consider perturbations that lower the symmetry of a
point defect, such as strain and electric field and how they affect
the polarization properties. We also show that the non-axial
spin-orbit interaction discussed in Ref. \cite{Tamarat:NJP2008} does 
not mix the eigenstates of the center in a given multiplet.
% as the spin-orbit interaction remain invariant under the symmetries of the $C_{3v}$ point group and the mixing it induces is suppressed by the large gap between the ground and excited state. 
%In particular, it should not mix the states of the lower branch of the excited state. 
Instead, we find that the electron spin-spin interaction is
responsible for the spin state mixing of the excited state as a result of
the lack of inversion symmetry of the center. 
%We also analyze possible sources of mixing between the states. 
Finally, we analyze the effect of electric fields via the inverse
piezoelectric effect and compare our results with experimental
observations. We show
that this effect can be used to tune the polarization properties of optical
transitions and the wavelength of emitted photons, which is of direct
importance for photon-based quantum communication between NV centers. Our
study clarifies important properties of NV centers and provides the
foundation for coherent interaction between electronic spins and photons in
solid state.

Our manuscript is organized as follows. In Section \ref{sec:states} we
present a general group theoretical formalism to calculate the
electronic or hole representation of a point defect for a given
crystal field symmetry and number of electrons contained in the
defect. Next, we use group theory and the explicit form of the states
to analyze the effect of the Coulomb interaction between electrons
(Section \ref{sec:ordering}) and spin-spin and spin-orbit interactions
for the NV center (Sections \ref{sec:spinspin} and
\ref{sec:spinorbit}, respectively). Next, we analyze the selection
rules of the unperturbed defect in Section
\ref{sec:selectionrules}. Finally, in Section \ref{sec:strain}, we
analyze the effect of strain and electric field perturbations.

\section{State representation}\label{sec:states}

We are particularly interested in quasi-static properties of defects
in crystals where the complex electronic structure can be observed
spectroscopically. In this limit one can apply the Born-Oppenheimer
approximation to separate the many-body system of electrons and
nuclei. This approximation relies on the fact that nuclei are much
slower than electrons. In this approximation the nuclei are
represented by their coordinates and the physical quantities of the
electrons depend on these coordinates as (external) fixed
parameters. A defect in a crystal breaks down the translational
symmetry reducing the symmetry of the crystal to rotations and
reflections. These symmetries form a point group which in general is a
subgroup of the point group of the lattice. The loss of translational
symmetry indicates that the Bloch-states are no longer a good
approximation to describe the point defect. In fact, some states can
be very well \emph{localized} near the point defect. These defect
states are particularly important in semiconductors and insulators
when they appear within the fundamental band gap of the crystal.

In the tight binding picture, the electron system of the diamond
crystal may be described as the sum of covalent-type interactions
between the valence electrons of two nearest neighbor atoms. When
defects involve vacancies, the absence of an ion will break bonds in
the crystal, producing unpaired electrons or dangling bonds,
$\sigma_i$, which to leading order can be used to represent the single
electron orbitals around the defect. The particular combination of
dangling bonds that form the single electron orbitals $\kp{\varphi_r}$
is set by the crystal field of the defect and can be readily
calculated by projecting the dangling bonds on each irreducible
representation (IR) of the point group of the
defect\cite{Tinkham:2003},
\begin{eqnarray}\label{eq:projection1}
\varphi_r = P^{(r)}\sigma_i = \frac{l_r}{h}\sum_e \chi^{(r)}_eR_e\sigma_i,
\end{eqnarray}
where $P^{(r)}$ is the projective operator to the IR $r$,
$\chi^{(r)}_e$ is the character of operation $R_e$ (element) for the
IR $r$, $l_r$ is the dimension of the IR $r$, and $h$ is the order of
the group (number of elements). A detailed application of
Eq. (\ref{eq:projection1}) for the case of the NV center can be found
in Appendix \ref{app:cha}. There are two non-degenerate totally
symmetric orbitals $a_1(1)$ ad $a_2(1)$ that transform according to
the one-dimensional IR $A_1$, and there are two degenerate states
$\kp{e_x, e_y}$ that transform according to the two-dimensional IR
$E$. At this stage, group theory does not predict the energy order of
these states. However a simple model of the electron-ion Coulomb
interaction can be used to qualitatively obtain the ordering of the
levels \cite{Lannoo:PRB1981}. In Appendix \ref{app:cha} we model the
effect of this interaction on the single electron orbitals,
$\varphi_r$, for the case of the NV center and find that the ordering
of the states (increasing in energy) is $a_1(1),a_1(2)$ and
$\kp{e_x,e_y}$.  Indeed, \emph{ab initio} density functional theory
(DFT) calculations revealed \cite{Goss:PRL1996,Gali:PRB2008} that the
$a_1(1)$ and $a_1(2)$ levels fall lower than the $e_x$ and $e_y$
levels, which demonstrates the strength of group theory for
\emph{qualitative} predictions.

Once the symmetry and degeneracy of the orbitals are determined, the
dynamics of the defect is set by the number of electrons available to
occupy the orbitals. The orbitals with higher energy will
predominantly set the properties of the defect. The spin character of
the defect will be determined by the degeneracy of the orbitals and
the number of electrons in them, leading to net spins $S=\kp{0, 1,
  2,...}$ if this number is even and $S=\kp{\frac{1}{2},
  \frac{3}{2},...}$ if odd.

In the case of the negatively charged NV center, each carbon atom
contributes one electron, the nitrogen (as a donor in diamond)
contributes two electrons, and an extra electron comes from the
environment \cite{Gali:PRB2008}, possibly given by substitutional
nitrogens \cite{Kok:Nature2006}. The ground state configuration
consists of four electrons occupying the totally symmetric states and
the remaining two electrons pairing up in the $\kp{e_{x},e_{y}}$
orbitals. In this single particle picture, the excited state
configuration can be approximated as one electron being promoted from
the $a_1(2)$ orbital to the $e_{x,y}$ orbitals \cite{Goss:PRL1996}.

If two more electrons were added to any of these configurations, the
wavefunction of the defect would be a singlet with a totally symmetric
spatial wavefunction, equivalent to the state of an atom with a filled
shell \cite{Stoneham:2001,Griffith:1961}. Therefore, the electronic
configuration of this defect can be modeled by two holes occupying the
orbitals $e_{x,y}$ in the ground state ($e^2$ electronic
configuration) and one hole each in the orbitals $a_1(2)$ and $e_{x,y}$ for
the excited state ($ae$ electronic configuration). A third electronic
configuration, $a^2$, can be envisioned by promoting the remaining
electron from the orbital $a_1(2)$ to the orbitals $e_{x,y}$. Hole and
electron representations are totally equivalent and it is convenient
to choose the representation containing the smallest number of
particles. If a hole representation is chosen, some care must be
taken, as some interactions reverse their sign, such as the spin-orbit
interaction \cite{Stoneham:2001}. In what follows, we choose a hole
representation containing two particles (instead of an electron
representation containing four particles), since it is more convenient
to describe the physics of the NV center. However, the analysis can be
applied to electrons as well.

The representation of the total $n$-electron wavefunction, including
space and spin degrees of freedom, is given by the direct product of
the representation of each hole $\Gamma_{hn}$ and its spin 
$\Gamma_\Psi = \prod_n
\rp{ \Gamma_{hn}\otimes D_{\frac{1}{2}}}$, where $D_{\frac{1}{2}}$ is
the representation for a spin $\frac{1}{2}$ particle in the
corresponding point group. 
%In the $C_{3v}$ point group, the spin representation can be reduced in terms of the irreducible representations $D_{\frac{1}{2}}\otimes D_{\frac{1}{2}} = A_1 + A_2 + E$, corresponding to the singlet ($A_1$) and triplet with zero ($A_2$) and non-zero ($E$) spin projections. We note that these representations belong to the IRs of the single $C_{3v}$ point group because of the two-particle system in our example. Nevertheless, the readers should not be confused with our labeling. Next, we label the states with both space and spin coordinates that are described by the double group of $C_{3v}$ point group. In odd-particle system, e.g. NV$^0$, one would obtain that IRs belong strictly to the double group of $C_{3v}$ point group.
The reduction or block diagonalization of the representation
$\Gamma_{\Psi}$ gives the eigenstates of the hamiltonian associated
with the crystal field potential and any interaction that remains
invariant under the elements of the point group in question. These
interactions include spin-orbit, spin-spin and Coulomb interactions,
as well as expansions, contractions, and stress where their axes
coincide with the symmetry axis of the defect. The eigenstates can be
found by projecting any combination of the two electron wavefunction
onto the irreducible representations of the group
\cite{Tinkham:2003,Jacobs:2005},
\begin{eqnarray}\label{eq:projection2}
\Psi^r = P^{(r)}\varphi_1\varphi_2 =
\frac{l_r}{h}\sum_e\chi^{(r)\ast}_e R_e\varphi_1 R_e\varphi_2,
\end{eqnarray}
where $\varphi_i$ can be any of the orbitals in Eq.
(\ref{eq:projection1}) and the subindex $i$ refers to the hole $i$. In
the case of the NV center, it is illustrative to note that the spin
representation for the two particles can be reduced to $D_{1/2}\otimes
D_{1/2} = A_1 + A_2 + E$, where $A_1$ corresponds to the singlet
state, and $A_2$ and $E$ to the triplet state with zero and non-zero
spin projections, respectively. A list of the eigenstates and their
symmetries for the two hole representation can be found in Table
\ref{t:states} for the ground state ($e^2$) and the excited state
($ae$). For completeness, we include the doubly excited state ($a^2$)
electronic configuration although this state is not optically
accessible in the excitation process of the NV center in
experiments. Note that each electronic configuration might have
singlet and triplet states. The calculation performed to obtain Table
\ref{t:states} is similar to the calculation made to find the
eigenstates when two spin particles are considered. However, in this
case one should use the Wigner coefficients of the corresponding
irreducible representation of the point group under consideration.

Group theory can predict why the hyperfine interaction with the
nuclear spin of the nitrogen in the excited state is more than an
order of magnitude larger than in the ground state for both nitrogen
species: the non-zero spin density in the ground state wavefunction of
the NV center is mostly concentrated in the orbitals $e_{x,y}$, which
have no overlap with the nitrogen atom. On the other hand, in the
excited state, when one electron is promoted from the $a_1(2)$ orbital
to one of the $e_{x,y}$ orbitals, the non-zero spin density comes now
from unpaired electrons occupying the orbitals $a_1(2)$ and
$e_{x,y}$. As the orbital $a_1(2)$ is partially localized on the
nitrogen atom, a sizable contact term interaction between the
electronic spin and the nuclear spin of the nitrogen is expected
\cite{Gali:PRB2008,Fuchs:PRL2008,Gali:PRB2009}.

Up to now, eigenstates inside a given electronic configuration have
the same energy, but the inclusion of the electron-electron Coulomb
interaction will lift the degeneracy between triplets and
singlets. The resulting energy splitting can be of the order of a
fraction of an eV and it is analyzed for the ground state
configuration of the NV center in Section
\ref{sec:ordering}. Furthermore, the degeneracy of triplet states is
lifted by spin-orbit and spin-spin interactions of the order of GHz,
where the crystal field plays an important role. These interactions
will be treated in sections \ref{sec:spinorbit} and
\ref{sec:spinspin}.

\begin{table}[h]
\caption{Partner functions of each IR for the direct product of two
  holes. The first column shows the electronic configuration and
  in parenthesis their triplet (T) or singlet (S) character. The
  last column shows the name of the state given in this paper and their
  symmetry. $\alpha$($\beta$) stands for
  $\uparrow$($\downarrow$) and $E_{\pm} = \ket{ae_{\pm}-e_{\pm}a}$,
  where $e_\pm = \mp(e_x\pm ie_y)$, $\ket{X} = (\ket{E_-} -
  \ket{E_+})/2$ and $\ket{Y} = (\ket{E_-} + \ket{E_+})i/2$.}

\begin{center}
\begin{tabular}{c|c|c@{}c} 
  \hline
  Conf. & State & \multicolumn{2}{c}{Name}\\
  \hline
\multirow{3}{*}{$e^2$ (T)}  
&\multirow{3}{*}{$\ket{e_xe_y-e_ye_x}\otimes\left\{ \begin{tabular}{c}
$|\beta\beta\rangle$ \\ $|\alpha\beta+\beta\alpha\rangle$ \\ $|\alpha\alpha\rangle$\end{tabular}\right.$}& $^3A_{2-}$ & ($E_1$) \\
&   & $^3A_{20}$ & ($A_1$) \\
&    & $^3A_{2+}$ & ($E_2$) \\
\multirow{3}{*}{$e^2$ (S)}  
&  \multirow{3}{*}{$\left.\begin{tabular}{c}
$\ket{e_xe_x-e_ye_y}$ \\ $\ket{e_xe_y+e_ye_x}$ \\ $\ket{e_xe_x+e_ye_y}$\end{tabular} \right\}\otimes\ket{\alpha\beta-\beta\alpha}$} & $^1E_1$  & ($E_1$)  \\
&   & $^1E_2$  & ($E_2$)  \\
&   & $^1A_1$  & ($A_1$) \\
  \hline
\multirow{6}{*}{$ea$ (T)}  
& $\ket{E_-}\otimes\ket{\alpha\alpha}-\ket{E_+}\otimes\ket{\beta\beta}$ & $A_1$ & ($A_1$) \\
&  $\ket{E_-}\otimes\ket{\alpha\alpha}+\ket{E_+}\otimes\ket{\beta\beta}$ & $A_2$ & ($A_2$) \\
&  $\ket{E_-}\otimes\ket{\beta\beta} - \ket{E_+}\otimes\ket{\alpha\alpha}$ & $E_1$ & ($E_1$) \\
&  $\ket{E_-}\otimes\ket{\beta\beta} + \ket{E_+}\otimes\ket{\alpha\alpha}$ & $E_2$ & ($E_2$) \\
& $\ket{Y}\otimes\ket{\alpha\beta+\beta\alpha}$ & $E_y$ & ($E_1$) \\
&  $\ket{X}\otimes\ket{\alpha\beta+\beta\alpha}$ & $E_x$ & ($E_2$) \\
  \hline
\multirow{2}{*}{$ea$ (S)}  
&  $\ket{a_1x+xa_1}\otimes\ket{\alpha\beta-\beta\alpha}$ & $^1E_x$ & ($E_1$) \\
&  $\ket{a_1y+ya_1}\otimes\ket{\alpha\beta-\beta\alpha}$ & $^1E_y$ & ($E_2$) \\
  \hline
\multirow{1}{*}{$a^2$ (S)}  
&  $\ket{a_1a_1}\otimes\ket{\alpha\beta-\beta\alpha}$ & $^1A_1$ & ($A_1$) \\
  \hline
\end{tabular}
\end{center}
\label{t:states}
\end{table}%

\section{Ordering of singlet states}\label{sec:ordering}

For a given electronic configuration, the most relevant interaction is
the electron-electron Coulomb interaction, which is minimized when
electrons are configured in an antisymmetric spatial configuration. As
the total wavefunction must be antisymmetric for fermionic particles,
the spin configuration must be symmetric. As a result, the state with
the largest multiplicity lies lower in energy. This analysis, known as
the first Hund's rule, predicts that the ground state of the NV
center should be the triplet $^3A_2$ state. We now address the
question related to the order of singlets in the ground state
electronic configuration $e^2$. The order of singlet states has a
great significance in understanding the spin-flipping fluoresence of
the NV center, and \emph{ab initio} DFT calculations were unable to
address this issue properly due to the many-body singlet states. Since
we have the explicit form of the wavefunctions, we can work out the
ordering of the singlets in a given electronic configuration by
analyzing the expectation value of the Coulomb interaction, which can
be written in the general form,
\begin{eqnarray}
C_{abcd} = \int dV_1dV_2 a^\star(r_1)b^\star\rp{r_2} V\rp{|r_1-r_2|}
c\rp{r_1}d\rp{r_2} \nonumber.
\end{eqnarray}
Using this expression we find that in the ground state electronic
configuration ($e^2$), the Coulomb interactions for these states are
\begin{eqnarray}\label{eq:coulomb}
C({^3A}_2) &=& \rp{ C_{xyxy} - C_{xyyx} - C_{yxxy} + C_{yxyx} }/2\nonumber\\
C({^1E}_1) &=& \rp{ C_{xyxy} + C_{xyyx} + C_{yxxy} + C_{yxyx} }/2\nonumber\\
C({^1E}_2) &=& \rp{ C_{xxxx} - C_{xxyy} - C_{yyxx} + C_{yyyy} }/2\nonumber\\
C({^1A}_1) &=& \rp{ C_{xxxx} + C_{xxyy} + C_{yyxx} + C_{yyyy} }/2  ,
\end{eqnarray}
where ${x,y}$ correspond to ${e_x,e_y}$ states.  From this set of
equations we find that the spacing between the singlets ${^1A}_1$ and
${^1E}_2$ is equal to the spacing between the singlet ${^1E}_1$ and
the ground state ${^3A}_2$, i.e., $C({^1A}_1) - C({^1E}_2) =
C({^1E}_1) - C({^3A}_2) = C_{xxyy}+C_{yyxx}\equiv 2e$, where the
difference is the exchange energy. In addition, as $^1E_1$ and $^1E_2$
belong to the same IR $E$, it can be shown that $C({^1E}_2) =
C({^1E}_1)$ (see Appendix \ref{a:ordering}). Under this consideration,
the ordering of the states is $\kp{^3A_2, {^1E}, {^1A}_1}$ with
relative energies $\kp{0, 2e, 4e}$. It should be noted that, in this
case, the most symmetric state has higher energy since the Coulomb
interaction between two electrons is repulsive. This picture might be
modified by the following effect. Since the Coulomb interaction
transforms as the totally symmetric IR, the matrix elements between
states with the same symmetry are non-zero. The states $^1E(e^2)$ and
${^1E}(ae)$ can couple via the Coulomb interaction, increasing the gap
between them. A similar effect happens with the states $^1A_1(e^2)$
and $^1A_1(a^2)$.  In Eq. (\ref{eq:coulomb}) we did
not take into account the effect of the other electrons present in the
system. 
%This might be possible by dividing the Coulomb-term by the dielectric constant of the host diamond. In this crude estimation, the rest of the electrons and the ions would shield the bare Coulomb-interaction. This is still an oversimplification, since the dielectric function is truly space-dependent and can differ considerably from the dielectric constant. 
Nevertheless, our basic results here serve as a \emph{qualitative}
estimate for the energy of levels and provides useful insight into the
structure of the NV center. The results of a very recent calculations 
based on many-body perturbation
theory (MBPT) \cite{Ma:PRB2010} supports our conclusion.

\section{Spin-Orbit interaction}\label{sec:spinorbit}

In the previous section the electronic spin did not directly enter
into our considerations. For instance, the energy of the $m_S=0,\pm 1$
sublevels of the ${^3A}_2$ ground state would have exactly the same
energy.  However, if the electronic spin is taken into account, one
can infer from Table \ref{t:states} that in general the $m_S=0$ and
$m_S=\pm1$ projections transform as functions of different IRs. For
example, in the ground state ${^3A_2}$, the $m_S=0$ projection
transforms as the IR $A_1$, while the $m_S=\pm1$ projections transform
as the IR $E$. This implies that the projections do not share the same
eigenenergies of the system. The spin-spin and spin-orbit interactions
may result in splitting of these orbitally degenerate states.

The spin-orbit interaction lifts the degeneracy of multiplets that
have non-zero angular momentum, and is also responsible for
transitions between terms with different spin states
\cite{Stoneham:2001}. It is a relativistic effect due to the relative
motion between electrons and nuclei. In the reference frame of the
electron, the nuclear potential, $\phi$, produces a magnetic field
equal to $\nabla \phi \times {\mathbf v}/c^2$. In SI units, this
interaction is given by
\begin{eqnarray}\label{eq:so}
H_{SO} = \frac{1}{2} \frac{\hbar}{c^2m_e^2}\rp{{\nabla} V \times \mathbf{p}}
\cdot \rp{\frac{\mathbf{s}}{\hbar}},
\end{eqnarray}
where $V =e\phi$ is the nuclear potential energy, $m_e$ is the
electron mass and $\mathbf{p}$ is the momentum. The presence of the
crystal field breaks the rotational symmetry of this
interaction. Since $\phi$ is produced by the nuclear potential, it
transforms as the totally symmetric representation $A_1$, and
therefore $\nabla V = \rp{V_x,V_y,V_z}$ transforms as a vector, where
$V_i = \partial V/\partial x_i$. Since $\mathbf{p}$ also transforms as
a vector, it is possible to identify the IRs to which the orbital
operator components $\vec{O} = \nabla V \times \mathbf{p} =$
$\rp{V_yp_z - V_zp_y,V_zp_x-V_xp_z,V_xp_y-V_yp_x}$ belong. In
$C_{3v}$, the components of $\nabla V$ and $\mathbf{p}$ transform as
$\rp{E_1,E_2,A_1}$ and therefore $\vec{O}$ transforms as the IRs
$\rp{E_2,E_1,A_2} = \rp{E,A_2}$.  The non-zero matrix elements of the
orbital operators $O_i$ in the basis $\kp{a, e_x, e_y}$ can be
determined by checking if $\rp{\varphi_i, O_k, \varphi_f} \supset A_1$
and are shown in Table \ref{t:spinorbit} where $A = \ave{e_y|O_x|a}$
and $B = \ave{e_x|O_z|e_y}$ (for simplicity we denote by $a$ the
$a_1(2)$ orbital state ). In this case, the spin-orbit interaction can
be written in terms of the angular momentum operators $l_i$ and takes
the following form:
\begin{eqnarray}
H_{SO}  =\lambda_{xy} \rp{l_xs_x + l_ys_y} + \lambda_{z}l_zs_z,
\end{eqnarray}
where $\lambda_{x,y}$($\lambda_z$) denotes the non-axial (axial)
strength of the interaction. In a system with $T_d$ or spherical
symmetry, $A=B$ and the usual form ($\mathbf{S}\cdot \mathbf{L}$) of the spin-orbit
interaction is recovered. It is also useful to think about $e_{\pm}$
as $p_\pm$ orbitals and $a_1(2)$ as a $p_z$ orbital, where the angular
momentum operators satisfy $l_{\pm}a_1(2)\propto
e_{\pm}$ \cite{Yu:2005}.

\begin{figure}[h]
\begin{center}
\includegraphics[width=0.75\textwidth]{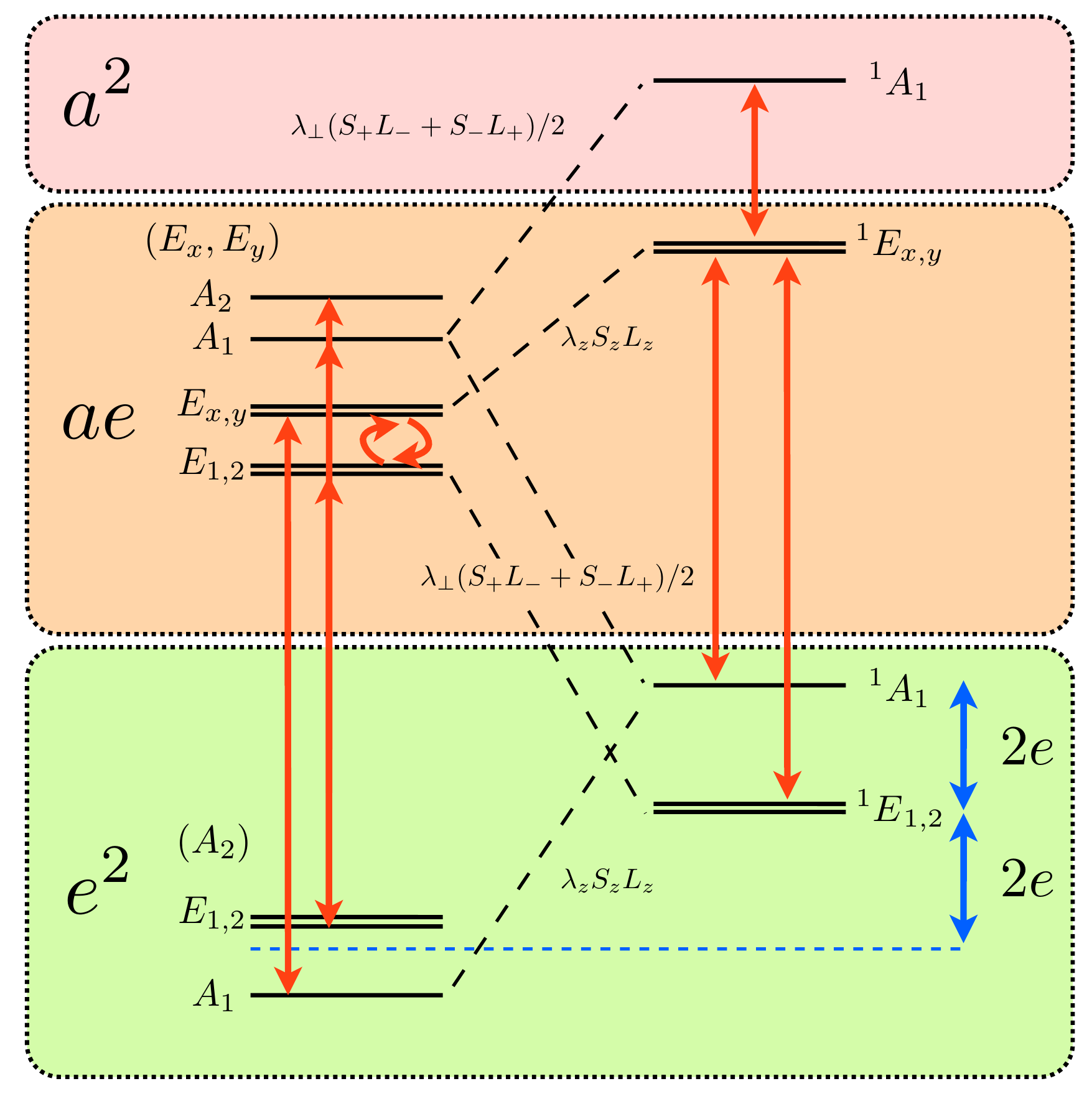}
\caption{{\bf Energy diagram of the unperturbed nitrogen-vacancy
    center in diamond.} Note that each electronic configuration can
  contain triplets (left column) as well as singlets (right column) which have been drawn in separated columns for clarity. Red arrows indicate allowed
  optical transitions via electric dipole moment interactions. The circular arrows between the states $E_{1,2}$ and $E_{x,y}$ represent the mixing due to spin-spin interaction (see Figure \ref{fig:soandss}). Dashed lines
  indicate possible non-radiate processes assisted by spin-orbit
  interaction. In the ground state ($e^2$ configuration), the distance
  between singlets and triplets is equal to the exchange energy of
  Coulomb interaction ($2e$). The horizontal dashed blue line represents the orbital energy of the ground state (without including spin-spin interaction).}
\label{fig:levels}
\end{center}
\end{figure}

\begin{table}[htdp]
\caption{Matrix elements for orbital operators in the $C_{3v}$ point
  group. For the $T_d$ symmetry group or spherically symmetric
  potentials, $A = B$.}
\begin{center}
\begin{tabular}{c|ccc}
$O_x$ & $\ket{e_x}$ & $\ket{e_y}$ & $\ket{a}$ \\
\hline
$\bra{e_x}$ & 0 & $0$ & $0$ \\
$\bra{e_y}$ & $0$ & 0 & $iA$ \\
$\bra{a}$ & $0$ & $-iA$ & 0 
\end{tabular}
\qquad
\begin{tabular}{c|ccc}
$O_y$ & $\ket{e_x}$ & $\ket{e_y}$ & $\ket{a}$ \\
\hline
$\bra{e_x}$ & $0$ & 0 & $-iA$ \\
$\bra{e_y}$ & 0 & $ 0 $ & $0$ \\
$\bra{a}$ & $iA$ & $0$ & 0 
\end{tabular}
\qquad
\begin{tabular}{c|ccc}
$O_z$ & $\ket{e_x}$ & $\ket{e_y}$ & $\ket{a}$ \\
\hline
$\bra{e_x}$ & 0 & $iB$ & 0 \\
$\bra{e_y}$ & $-iB$ & 0 & 0 \\
$\bra{a}$ & 0 & 0 & 0 
\end{tabular}
\end{center}
\label{t:spinorbit}
\end{table}%
Once it is known how the spin-orbit interaction acts on the orbitals,
$e_x, e_y$ and $a$, it is possible to calculate the effect of this
interaction on the 15 states given in Table~\ref{t:states}. An
important effect is the splitting in the excited state triplet between
the states $A_1, A_2$ and $E_x, E_y$ and between states $E_x, E_y$ and
$E_1, E_2$ \cite{Lenef:PRB1996}. The spin-orbit interaction can be written as,
\begin{eqnarray}\label{eq:soae}
H_{SO} = \lambda_z(\ketbra{A_1}{A_1} + \ketbra{A_2}{A_2} - \ketbra{E_1}{E_1} - \ketbra{E_2}{E_2}),
\end{eqnarray}
in the excited state triplet manifold $\{A_1,A_2,E_x,E_y,E_1,E_2\}$. 
Another effect, relevant when
treating non-radiative transitions, is that the axial part of the
spin-orbit interaction ($\lambda_z$) links states with $m_s=0$ spin
projections among states of the same electronic configuration, while
the non-axial part ($\lambda_{x,y}$) links states with non-zero spin
projections with singlets among different electronic
configurations. In Figure~\ref{fig:levels} we show the states linked
by the axial and the non-axial parts of the spin-orbit interaction,
for which non-radiative transitions might occur. In addition to the
well known transition between $A_1(ae)\rightarrow {^1A}_1(e^2)$, we
find that this interaction might also link $E_{1,2}(ae)\rightarrow
{^1E_{1,2}}(e^2)$ and in particular $E_{x,y}\rightarrow
{^1E_{x,y}}(ae)$. The latter transition may play an important role, as
recent \emph{ab initio} calculations have shown that the singlets
${^1E_{x,y}}$ might lie very close in energy to the excited state
triplet \cite{Ma:PRB2010}. In our model, the non-axial part of the
spin-orbit interaction, $\lambda_{x,y}\rp{l_+s_-+l_-s_+}$, does not
mix the states of the excited state triplet with different spin
projections because the raising and lower operators, $l_-$ and $l_+$,
link states of different electronic configurations. In particular,
this interaction cannot mix the states of the excited state triplet
because the mixing is suppressed by the large energy gap that
separates different electronic configurations.

We have numerically evaluated the ratio between the axial part and
transverse part of spin-orbit, $\lambda_z/\lambda_{xy} = B/A = 0.75$
using the functions $e_x$ and $e_y$ and $a_1(2)$ from \emph{ab initio}
calculations (see Appendix \ref{app:compmethods}). This suggest that
if the axial part of spin-orbit is 5.5 GHz \cite{Batalov:PRL2009}, the
non-axial part should be on the order of $\lambda_{xy} = 7.3$ GHz and
only couples singlets with triplets states as shown in
Figure\ref{fig:levels}. We have also numerically confirmed the
structure of Table \ref{t:spinorbit} with three digits of precision in
units of GHz (see Appendix \ref{app:compmethods}).

\section{Spin-spin interaction}\label{sec:spinspin}

The spin-spin interaction between electrons is usually not present in
systems with spherical symmetry, due to the traceless character of the
magnetic dipole-dipole interaction. However, if the electron
wavefunction is not spherically distributed, this interaction does not
average out. Here we describe its effect on the excited state triplet
of the NV center and we provide a numerical estimation of its
strength. The spin-spin interaction can be written (in SI units) as,
\begin{eqnarray}\label{eq:spinspin}
h_{ss} &=&
-\frac{\mu_0}{4\pi}\frac{g^2\beta^2}{r^3}\rp{3(\mathbf{s}_1\cdot
  \hat{r})(\mathbf{s}_2\cdot \hat{r}) - \mathbf{s}_1\cdot
  \mathbf{s}_2},
\end{eqnarray}
where $\mathbf{s}_i = \frac{1}{2}\sqp{\sigma_x, \sigma_y, \sigma_z}$
are the spin operators of particle $i$ and $\sigma_j$ $(j=x,y,z)$ are the Pauli
matrices, $\beta$ is the Bohr magneton, $g$ is the Land\'{e}-factor
for the electron and $\mu_0$ is the magnetic permeability of free
space \cite{note1}
%\footnote{The contact term does not contribute due to the Pauli
%  exclusion principle.}
. In order to analyze the effect of this
interaction in the defect it is useful to write the spatial and spin
parts separately in terms of the irreducible representations of the
point group. Then, it is straightforward to express this interaction
in terms of the eigenstates of the defect (see Appendix
\ref{app:spinspin}),
\begin{eqnarray}\label{eq:nvspinspin}
H_{ss} &=& \Delta\left( \ketbra{A_1}{A_1} + \ketbra{A_2}{A_2} +
\ketbra{E_1}{E_1} + \ketbra{E_2}{E_2} \right)\nonumber \\&&- 2\Delta
\left( \ketbra{E_x}{E_x} + \ketbra{E_y}{E_y} \right) \nonumber \\&&+
2\Delta'\left( \ketbra{A_2}{A_2} - \ketbra{A_1}{A_1} \right) \nonumber
\\&&\Delta''\rp{\ketbra{E_1}{E_y}+\ketbra{E_y}{E_1}-i
  \ketbra{E_2}{E_x}+i\ketbra{E_x}{E_2}} ,
\end{eqnarray}
where the gaps between the $m_s=\pm1$ and $m_s=0$ projections and
between $A_1$ and $A_2$ states are given by
\begin{eqnarray}
3\Delta &=& 3\frac{\mu_0}{4\pi}g^2\beta^2\left\langle X\left|\frac{1 -
  3\hat{z}^2}{4r^3}\right| X \right\rangle =
-\frac{3}{4}D_{zz}\label{eq:ssparam1} \\ 4\Delta^\prime &=&
4\frac{\mu_0}{4\pi}g^2\beta^2\left\langle X\left|
\frac{3\hat{x}^2-3\hat{y}^2}{4r^3}\right| X \right\rangle =
D_{x^2-y^2},\label{eq:ssparam2}
\end{eqnarray}
while the mixing term is given by
\begin{eqnarray}
\Delta^{\prime\prime} &=& \frac{\mu_0}{4\pi}g^2\beta^2\left\langle
X\left| \frac{3\hat{x}\hat{z}}{\sqrt{2}r^3}\right| X
\right\rangle. \label{eq:ssparam3}
\end{eqnarray}

\begin{figure}[h]
\begin{center}
\includegraphics[width=0.95\textwidth]{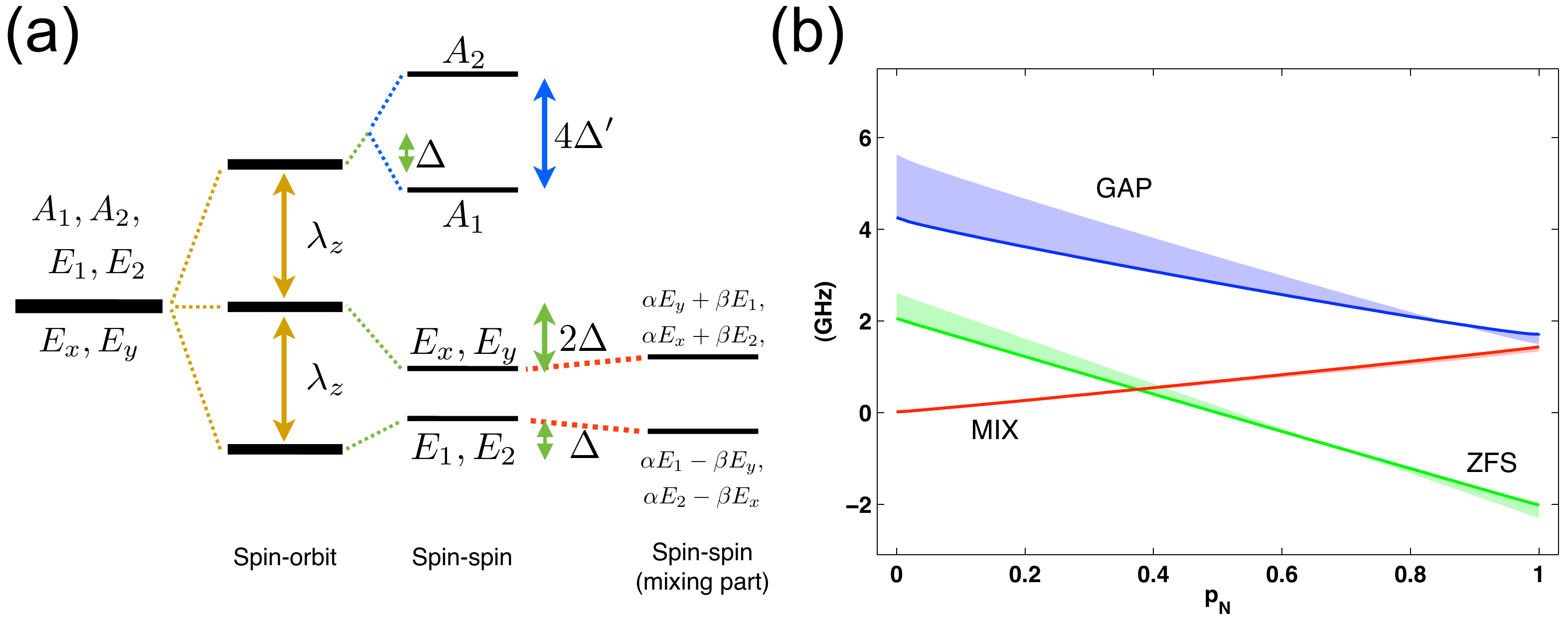}
\caption{{\bf Splitting due to spin-orbit and spin-spin interaction in
    triplet $ae$.} {\bf (a)} The axial part of the spin-orbit interaction splits the states $\{A_1,A_2\}$, $\{E_x,E_y\}$ and $\{E_1,E_2\}$ by $\lambda_z$. The spin-spin interaction splits states with different spin projections and also splits the $A_{1}$ and $A_2$ states. Our
  theory predicts the $A_2$ state at higher energy than the $A_1$ state and that the
  states $\rp{E_{1,2}}$ and $\rp{E_{x,y}}$ are mixed. As the state $A_1$
  has an additional non-radiative decay channel, it is possible to
  confirm this finding by measuring the lifetime of the state. Note
  that the splitting between $A_1$ and $A_2$ is a direct consequence
  of spin-orbit mixing the spatial and spin part of the
  wavefunction. {\bf (b)} Values for the zero field splitting
  ($3\Delta$), gap between the states $A_1$ and $A_2$ ($4\Delta'$) and
  mixing term ($\Delta''$) due to spin-spin interaction in the excited
  state as a function of the nitrogen population, $p_N$, in the state
  $a_1(2)$. The shadowed areas indicate the possible values for these
  parameters when the distance between the vacancy and the three
  carbons is increased between 0 and 3\%, and the distance between the
  vacancy and the nitrogen is decreased between 0 and 4\% of their
  excited state configuration. The solid lines correspond to the
  maximum (minimum) distance between the carbons (nitrogen) and the
  vacancy.}
\label{fig:soandss}
\end{center}
\end{figure}

Figure~\ref{fig:soandss} shows the effect of spin-orbit and spin-spin
interactions on the excited state manifold. In particular, we find
that the state $A_2$ has higher energy than the state $A_1$ ($2\Delta'>0$), contrary
to previous estimations \cite{Lenef:PRB1996,note2}
%\footnote{Recently, this
%  was indirectly experimentally confirmed. The lower energy state
%  $A_1$ was observed to have a shorter lifetime than the state
%  $A_2$\cite{Togan:Nature2010}. This is as expected since the state
%  $A_1$ decays non-radiatevely to the singlet $^1A_1$ via non-axial
%  spin-orbit.}
. In addition, we find that the spin-spin
interaction $\Delta''$ mixes states with different spin-projections. This effect
is the result of the lack of inversion symmetry of the NV center and
it is not present in systems with inversion symmetry such as free
atoms or substitutional atoms in cubic lattices. This does not contradicts group theoretical estimates as the mixed states transform
according to the same IR (e.g. the $E_1$ and $E_y$ states both transform
according to the IR $E_1$, see Table \ref{t:states}).

We estimated these parameters using a simplified model consisting of
the dangling bonds given in Figure \ref{fig:bonds} (in the Appendix)
for the three carbons and the nitrogen atom around the vacancy. The
dangling bonds are modeled by Gaussian orbitals that best fit to the
wavefunction obtained by an \emph{ab initio} DFT supercell calculation
(see Appendix \ref{app:compmethods}). The distance between atoms is
also taken from these simulations. To avoid numerical divergences when
$r=0$, we estimate Eq.  (\ref{eq:ssparam1}-\ref{eq:ssparam3}) in
reciprocal space following Ref. \cite{Rayson:PRB2008}. The values for
the zero field splitting ($\Delta_{es}=3\Delta$), gap between states
$A_1$ and $A_2$ ($4\Delta'$) and mixing term between states $E_{1,2}$
and $E_{x,y}$ ($\Delta''$) are given in Figure \ref{fig:soandss}b. As
{\it ab initio} calculations cannot accurately estimate the nitrogen
population $p_N = |\beta|^2$ in the single orbital state $a_1(2)$ (see
Appendix \ref{app:cha} for a definition of parameter $\beta$), we have
plotted in Figure \ref{fig:soandss}b, the values of the spin-spin
interaction as a function of $p_N$. In addition, the solid regions in
the figure take into account variations of the relative distance among
the three carbons, the nitrogen and the vacancy. The distance between
the carbons and the vacancy is increased between 0 and 3\%, meanwhile
the distance between the nitrogen and the vacancy is decrease between
0 and 4\% relative to their excited state configuration (solid
lines). This shows how the spin-spin interaction depend on the
distance between the atoms.

%We have evaluated numerically the values for this interaction using
%the orbitals $e_x$ and $e_y$ from \emph{ab initio} calculations (see
%Appendix \ref{app:compmethods}). To avoid numerical divergences when
%$r=0$, we estimate Eq.s (\ref{eq:ssparam1}) and (\ref{eq:ssparam2})
%in reciprocal space following reference \cite{Rayson:PRB2008}. We
%find that the zero field splitting on the ground state is 2.2 GHz
%compared to the 2.87 GHz found experimentally. Meanwhile, for the
%excited state, we find that $3\Delta = 0.2$ GHz and $4\Delta' = 2.5$
%GHz. This is compared with the experimental values 1.4 GHz and 3.3
%GHz measured in \cite{Batalov:PRL2009}.

We emphasize that, contrary to the ground state of the NV center, the
splitting between $A_1$ and $A_2$ in the excited state exists because
the spin-orbit interaction mixes the spin and spatial parts. In fact,
at high temperatures, where the spin-orbit interaction averages out
\cite{Rogers:NJP2009}, and if the spatial part is given by
$\ketbra{X}{X}+\ketbra{Y}{Y}$, it can be checked by looking at
Eq. (\ref{eq:spinspinxy}) that only the zero field splitting
$\Delta_{es}$ survives from the electronic spin-spin interaction, as
confirmed by experiments \cite{Fuchs:PRL2008,Rogers:NJP2009}. In addition, the spin-orbit interaction in the excited state, Eq. (\ref{eq:soae}), can be written as $H_{SO}=i(\ketbra{X}{Y}-\ketbra{Y}{X})\otimes(\ketbra{\alpha\alpha}{\alpha\alpha}-\ketbra{\beta\beta}{\beta\beta})$, which also vanishes if the spatial part is given by $\ketbra{X}{X}+\ketbra{Y}{Y}$.

\section{Selection rules and spin-photon entanglement schemes}
\label{sec:selectionrules}

%In our paper we simplify the excitation from the $^3A_2$ ground state
%to the $^3E$ excited state by promoting the electron from the single
%particle $a_1(2)$ level to the single particle $e_{x,y}$ levels. A
%recent \emph{ab initio} calculation could show \cite{Gali:PRL2009}
%that this simple picture holds very well for the NV center. The
%success of this simple treatment may be due to several factors such
%as i) $a_1(2)$ and $e_{x,y}$ levels fall deeply in the gap and are
%well-separated from the crystalline bands, ii) the ground and excited
%states belong to different IRs and there is no other triplet states
%in this energy range at all, so they do not mix with other states
%upon excitation. The excitation and fluorescence of NV center can be
%studied in this simplified picture. Apparently, the excitation causes
%significant change in the dipole moment
%\cite{Gali:PRB2009}. Understanding the selection rules is crucial to
%analyze absorption and fluorescence experiments in detail. As the
%electric dipole part of radiation only acts on the spatial part of
%the wavefunction, it can only make transitions between states that
%share the same spin.

Group theory tells that transitions are dipole allowed 
if the matrix element contains the totally symmetric IR,
$\braket{\varphi_f|\hat{e}}{\varphi_i}\supset A_1$. In the case of the
NV center ($C_{3v}$), the only non-zero matrix elements are
$\bra{a}\hat{x}\cdot r\ket{e_x}$ and $\bra{a}\hat{y}\cdot r\ket{e_y}$,
from which it is straighforward to calculate the selection rules among
the 15 eigenstates given in Table~\ref{t:states} for the unperturbed
center. This is shown in Table~\ref{t:rules}. These matrix elements
have been confirmed by our first-principles calculations of these
matrix elements in the velocity representation as well as by other
authors only for the triplet transition
\cite{Hossain:PRL2008}. %Interestingly, since the excited state triplet has non-zero angular momentum, photons emitted to the ground state must have left or right circular polarization in order to preserve total angular momentum. 
In addition to the well known triplet-triplet
transition \cite{Reddy:JL1987}, transitions are allowed between
singlets of different electronic configurations. We remark that the
transition between singlet $^1A_1(e^2)$ and singlet $^1E(e^2)$ is not
strictly forbidden by group theory to first order, but since both
states belong to the same electronic configuration, no dipole moment
exists between them and the probability of radiative transition is
extremely low. According to our results using wave functions from first-principles calculations (see Appendix
\ref{app:compmethods}), the ratio between the dipole transition matrix elements
associated with the singlet states to those of the triplet states is
about 5$\times$10$^{-9}$. The singlet-singlet transition might be allowed by phonons or
mixing of the states with singlets of different electronic
configurations. Recent experiments by Rogers \emph{et al.} identified an emission from singlet to
singlet\cite{Rogers:NJP2009}, which we suggest is related to the
$^1E(ae)\rightarrow^1A_1(e^2)$ transition. The transition
$^1A_1(e^2)\rightarrow^1E(e^2)$ might be possible for the reasons
described above, but it is unlikely to be sizable. A recent MBPT
calculation supports our conclusion \cite{Ma:PRB2010}. A suitable
experiment to unravel this issue would be to look at the presence of
this emission under resonant excitation. In this case, if the state
$^1E(ae)$ is above the excited state triplet, the state $^1E(ae)$ will
be hardly populated and therefore no singlet-singlet transition should
be observed.

\begin{figure}[h]
\begin{center}
\includegraphics[width=0.5\textwidth]{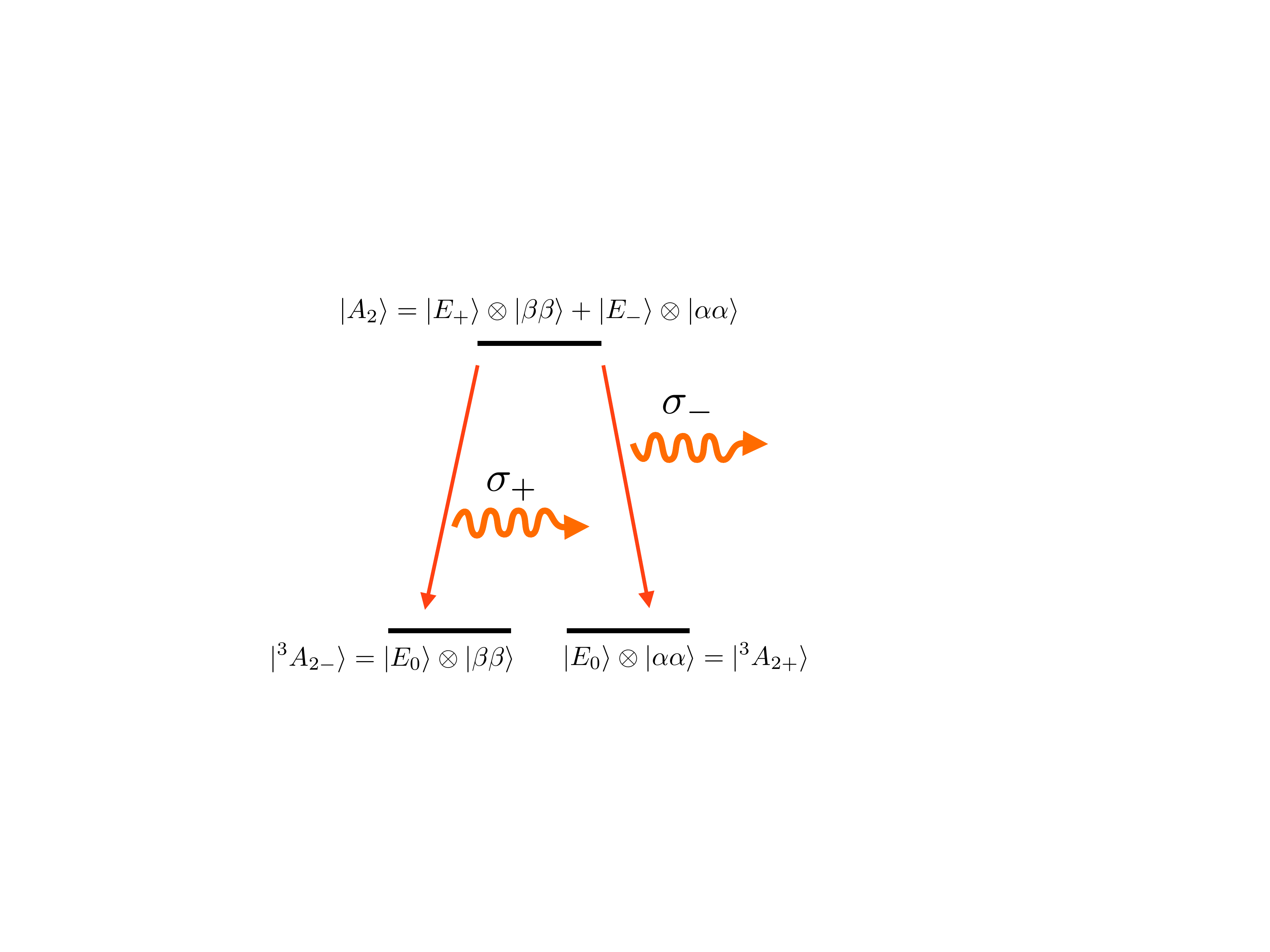}
\caption{{\bf Spin-photon entanglement generation.} When the NV center
  is prepared in the excited state $A_2({^3E})$, the electron can
  decay to the ground state ${^3A_2}$ $m_s=1$ ($m_s=-1$) by emitting a
  right (left) circularly polarized photon.}
\label{fig:entanglement}
\end{center}
\end{figure}

Once the selection rules are known for the defect, it is possible to
realize interesting applications such as spin-photon entanglement
generation \cite{Blinov:Nature2004}. In the case of the NV center, the
system can be prepared in the $A_2(ae)$ state. Next, the electron can spontaneously decay to the ground state ${^3A_{2-}}$ by
emitting a photon with $\sigma_{+}$ (right circular) polarization or to
the state ${^3A_{2+}}$ by emitting a $\sigma_{-}$ polarized photon
(see Figure \ref{fig:entanglement}). As a result, the spin of the
electron is entangled with the polarization (spin) of the photon. The
implementation of this scheme is sensitive to strain, which will be
analyzed in Section \ref{sec:strain}. However, in Section
\ref{sec:strainelectric}, we recognize that the application of an
electric field can be used to overcome some of these issues and
facilitate the next step of entangling between two NV centers.

\begin{table}[htdp]
\caption{Selection rules for optical transitions between: the triplet
  excited state $(ae)$ and the triplet ground state $(e^2)$, the
  singlets $(ae)$ and the singlets $(e^2)$, and the singlet $(a^2)$
  and the singlets $(ae)$. Linear polarizations are represented by
  $\hat{x}$ and $\hat{y}$, while circular polarizations are
  represented by $\hat{\sigma}_\pm = \hat{x}\pm i\hat{y}$. As an
  example, a photon with $\sigma_+$ polarization is emitted when the
  electron decays from state $A_2(ae)$ to state $^3A_{2-}(e^2)$. }
\begin{center}
\begin{minipage}[c]{0.45\linewidth}
\begin{tabular}{c|cccccc}
$\hat{e}$ &  $A_{1}$ & $A_{2}$ & $E_{1}$ & $E_{2}$ & $E_{x}$ & $E_{y}$ \\ \hline
$^3A_{2-}$ &  $\hat{\sigma}_+$ & $\hat{\sigma}_+$ & $\hat{\sigma}_-$ & $\hat{\sigma}_-$  & & \\
$^3A_{20}$ & & & & & $\hat{y}$ & $\hat{x}$  \\
$^3A_{2+}$ &  $\hat{\sigma}_-$ & $\hat{\sigma}_-$ & $\hat{\sigma}_+$ & $\hat{\sigma}_+$  & &
\end{tabular}
\end{minipage}
\quad
\begin{minipage}[c]{0.25\linewidth}
\begin{tabular}{c|cc}
$\hat{e}$ &  $^1E_{x}$ & $^1E_{y}$  \\ \hline
$^1A_{1}$ &  $\hat{x}$ & $\hat{y}$ \\
$^1E_{1}$ & $\hat{x}$ & $\hat{y}$ \\
$^1E_{2}$ &  $\hat{y}$ & $\hat{x}$
\end{tabular}
\end{minipage}
\quad
\begin{minipage}[c]{0.2\linewidth}
\begin{tabular}{c|c}
$\hat{e}$ &  $^1A_{1}$ \\ \hline
$^1E_{1}$ & $\hat{x}$ \\
$^1E_{2}$ &  $\hat{y}$
\end{tabular}
\end{minipage}
\end{center}
\label{t:rules}
\end{table}%

\section{The effect of strain}\label{sec:strain}

Strain refers to the displacement $\Delta u$ of the atomic positions
when the crystal is stretched stretch $\Delta x$ \cite{Yu:2005}. It is
a dimensionless tensor expressing the fractional change under
stretching, $e_{ij} = \frac{\partial{\delta{R_i}}}{\partial x_j}$, and
it can be produced by stress (forces applied to the solid structure),
electric field, or temperature \cite{Nye:1985}. A systematic study of
strain can be used to unravel the symmetry of defects and explore
their properties \cite{Davies:PRSLA1976}. Strain can shift the energy
of the states as well as mix them. It can reduce the symmetry of the
crystal field by displacing the atoms. However, not all nine
components of strain change the defect in a noticeable way. The
antisymmetric part of $e_{ij}$ transforms as a generator of the
rotational group and therefore only rotates the whole structure. The
symmetry and energies of the unperturbed states do not change upon
rotation. Only the symmetric part of strain, $\epsilon = e+e^T$ affect
the structure of a defect \cite{Yu:2005}. As with any other element of
the theory, strain can be expressed in terms of matrices that
transform according to the IRs of the point group under
consideration. These matrices can be found by projecting a general
strain matrix on each IR,
\begin{equation}
\epsilon_r = \frac{l_r}{h}\sum_e \chi_e^{\ast} R^{\dagger}_e\epsilon R_e.
\end{equation}
In Appendix \ref{app:strain} we show in detail how to deteermine the effect
of strain on the eigenstates of the defect. For simplicity, in the
case of the NV center we only write the effect of strain in the
manifold $\kp{e_x,e_y, a}$,
\begin{eqnarray}
H_{strain} = \delta^a_{A1}A_1^a + \delta_{A1}^bA_1^b + \delta_{E1}^aE_1^a + \delta_{E2}^aE_2^b + \delta_{E1}^aE_{1}^a + \delta_{E2}^bE_2^b
\end{eqnarray}
where $\delta^a_{A1} = (e_{xx}+e_{yy})/2$, $\delta^b_{A1}=e_{zz}$,
$\delta_{E1}^a = (e_{xx}-e_{yy})/2$,
$\delta_{E2}^a=(e_{xy}+e_{yx})/2$, $\delta_{E1}^b =
(e_{xz}+e_{zx})/2$, $\delta_{E2}^b=(e_{yz}+e_{zy})/2$ and
\begin{eqnarray}
{A_{1}^a} &=& {\tiny \matrixtt{1}{0}{0}{0}{1}{0}{0}{0}{0}}\qquad
{E_{1}^a} = {\tiny \matrixtt{1}{0}{0}{0}{-1}{0}{0}{0}{0}}\qquad
{E_{2}^a} = {\tiny \matrixtt{0}{1}{0}{1}{0}{0}{0}{0}{0}}\qquad\\\nonumber
{A_{1}^b} &=& {\tiny \matrixtt{0}{0}{0}{0}{0}{0}{0}{0}{1}}\qquad
{E_{1}^b} = {\tiny \matrixtt{0}{0}{1}{0}{0}{0}{1}{0}{0}}\qquad
{E_{2}^b} = {\tiny \matrixtt{0}{0}{0}{0}{0}{1}{0}{1}{0}},
\end{eqnarray}
in the manifold $\kp{e_x,e_y,a}$. The effect of strain on the orbitals
$a, e_x, e_y$ is easy to see.  ${A_{1}^a}$ will shift equally the
energies of the states $e_x$ and $e_y$, while ${A_{1}^b}$ will shift
the energy of states $a$. Note that both describe axial stress: the
former leaves the $e^2$ electronic configuration unaffected and the
latter leaves the $ea$ configuration unaffected. Either one produces
relative shifts between both configurations, resulting in an
inhomogeneous broadening of the optical transitions. However, they do
not change the selection rules. Only the stress ${A_{1}^a}+{A_{1}^b}$,
corresponding to either expansion or contraction, leaves all relative
energies unaffected. ${E_{x}^a}$ splits the energy between $e_x$ and
$e_y$ and ${E_{y}^a}$ mixes the two states. Finally, ${E_{x}^b}$ and
${E_{y}^b}$ mixes the states $e_x$ and $a_1$ and $e_y$ and $a_1$,
respectively. In the case of the NV center, the effect of the matrices
$E^b_{x,y}$ can be neglected thanks to the large gap between orbitals
$a$ and $e_{x,y}$. Therefore, in what follows we do not consider them
further.

Recent work has been done to analyze how strain affects the excited
state structure of the NV center
\cite{Rogers:NJP2009,Batalov:PRL2009}. Here we derive the explicit
form of strain affecting the different electronic configurations and
look at how strain affects the selection rules described in Section
\ref{sec:selectionrules}.

\begin{figure}[h]
\begin{center}
\includegraphics[width=0.85\textwidth]{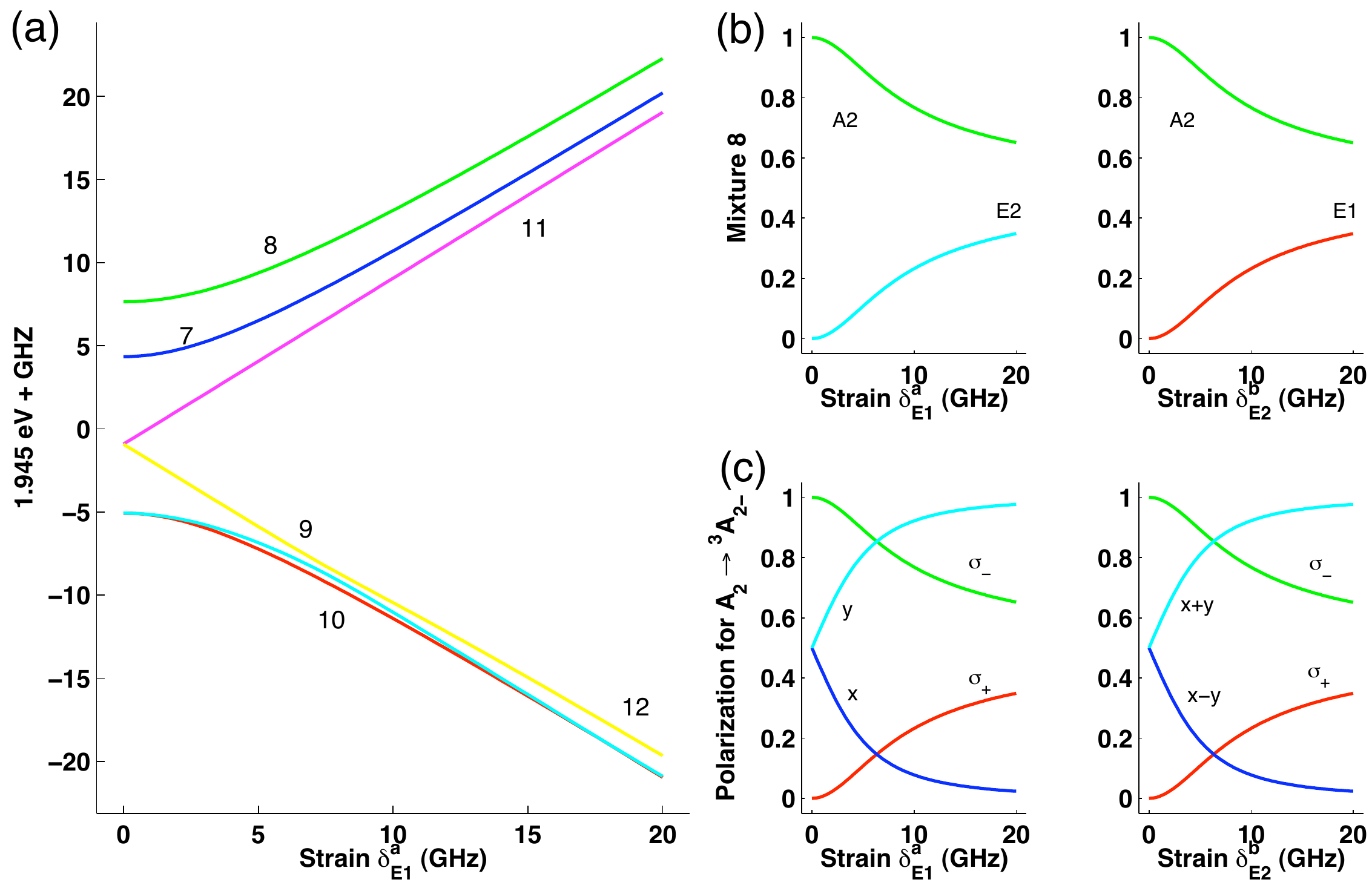}
\caption{{\bf Excited state structure as a function of strain.} (a)
  Eigenvalues of the excited state triplet as a function of
  $\delta_{E1}^a$ strain. (b) Mixture of the eigenstate with higher
  energy (corresponding to $A_2$ in the limit of low strain) and (c)
  the polarization of dipolar radiation under transitions from this
  state to the $^3A_{2+}$ state of the ground state. Note that in both
  cases the circular polarization character of radiation remains. On
  the other hand, the linear polarization rotates 90$^\circ$ for
  strain along $\delta_{E2}^b$ with respect to that of strain along
  $\delta_{E1}^a$.}
\label{fig:polarization}
\end{center}
\end{figure}

The relevant strain matrices we will consider are $E^a_x$ and $E^a_y$,
for which the Hamiltonian is,
\begin{eqnarray}\label{eq:strain}
H_{strain} = \delta_{E1}^a\rp{\ketbra{e_x}{e_x} - \ketbra{e_y}{e_y}} +
\delta_{E2}^b \rp{\ketbra{e_x}{e_y}+\ketbra{e_y}{e_x}}.
\end{eqnarray}
This mostly affects the singlet and excited state configurations in
the following form,
\begin{eqnarray}
\left( \begin{tabular}{cc|cc|cc}
  &  &  &  & $\delta_{E1}^a$ & $-i\delta_{E2}^b$ \\ 
 &   & &  & $-i\delta_{E2}^b$ & $\delta_{E1}^a$ \\ \hline
 &  & $\delta_{E1}^a$ & $\delta_{E2}^b$  & & $$ \\
 & & $\delta_{E2}^b$ & $-\delta_{E1}^a$  & $$ & \\ \hline
$\delta_{E1}^a$ & $i\delta_{E2}^b$ & & $$ &  & \\ 
$i\delta_{E2}^b$ & $\delta_{E1}^a$ & $$ & & &  
\end{tabular}
\right)\quad
\rp{
\begin{tabular}{ccc}
 & & $2\delta_{E1}^a$ \\
 & & $2\delta_{E2}^b$ \\
$2\delta_{E1}^a$ & $2\delta_{E2}^b$ &
\end{tabular}
}
\rp{
\begin{tabular}{cc}
$\delta_{E1}^a$ & $\delta_{E2}^b$ \\
$\delta_{E2}^b$ & $-\delta_{E1}^a$ \\
\end{tabular}
},
\end{eqnarray}
for the manifolds $\kp{A_1, A_2, E_x, E_y, E_1, E_2}$,
$\kp{^1E_1,{^1E}_2, {^1A}_1}$ and $\kp{^1E_x, {^1E}_y}$, respectively.
The ground state, due to its antisymmetric combination between $e_x$
and $e_y$, is stable under the perturbation $H_{strain}$. This can be
checked by applying Eq. (\ref{eq:strain}) to the ground state given in
Table \ref{t:states}. The effect on the excited state triplet can be
seen in Figure~\ref{fig:polarization}a, where the unperturbed states
are mixed in such a way that, in the limit of high strain, the excited
triplet structure splits into two triplets with spatial wavefunctions
$E_x$ and $E_y$. When strain overcomes the spin-orbit interaction
($\delta_{E1}^a>5.5$ GHz), the spin part decouples from the spatial
part and the total angular momentum is no longer a good quantum
number. Transitions from the excited state triplet to the ground state
triplet are linearly polarized, where the polarization indicates the
direction of strain in the $xy$ plane.

Figure~\ref{fig:polarization}c shows how the polarization of the
emitted photon from the state $A_2$ to the ground state $^3A_{2-}$,
varies from circular to linear as a function of strain. In the case of
$\delta_{E2}^b$ strain, the effect is similar but now the mixing is
different. As shown in Figure~\ref{fig:polarization}, $A_2$ mixes with
$E_1$ and the photons become polarized along $x-y$. \emph{Note that,
  in the limit of low strain, in both cases the polarization remains
  right circularly polarized for the transition between the excited
  state $A_2(ae)$ to the ground state${^3A}_{2-}(e^2)$, while the
  polarization remains left circular for the transition between the
  excited state $A_2(ae)$ to the ground state ${^3A}_{2+}(e^2)$.} The
fact that at lower strain the character of the polarization remains
circular has been succesfully used in entanglement schemes
\cite{Togan:Nature2010}. The polarization properties of the states
$E_{1,2}$ are similar to those of the states $A_{1,2}$ but with the
opposite polarization.

%It is important to note that the effect of external strain depends on
%the gap between states. As an example, we analyze the gap between
%states of the lower branch of the excited state triplet under
%strain. On Figure~\ref{fig:gap}, we show the gap among the states
%$\kp{E_x, E_1, E_2}$ as a function of strain. Notice that at
%$\delta_{E1}^a = 7$~GHz the gap between states $E_x$ and $E_1$
%reduces to zero and the states cross each other. At this point any
%small perturbation that mixes the states can transform this crossing
%into an anticrossing point. In particular, we have shown that the
%spin-spin interaction can mix the states and also the addition of an
%external magnetic field or the interaction with the nuclear spin of
%the nitrogen \cite{Gonzalez:PRB2009} can cause this effect. For low
%magnetic field, the interaction the most contribute to the mixing of
%the states and, therefore, to non-spin-preserving transitions is the
%spin-spin interaction. This effect is enhanced when strain reduces
%the gap between states. We emphasize that the non-axial spin-orbit
%interaction cannot cause this mixing as it is invariant under
%$C_{3v}$ and it only mixes states from different electronic
%configurations.

\section{Strain and Electric field}\label{sec:strainelectric}

\begin{figure}[h]
\begin{center}
\includegraphics[width=0.75\textwidth]{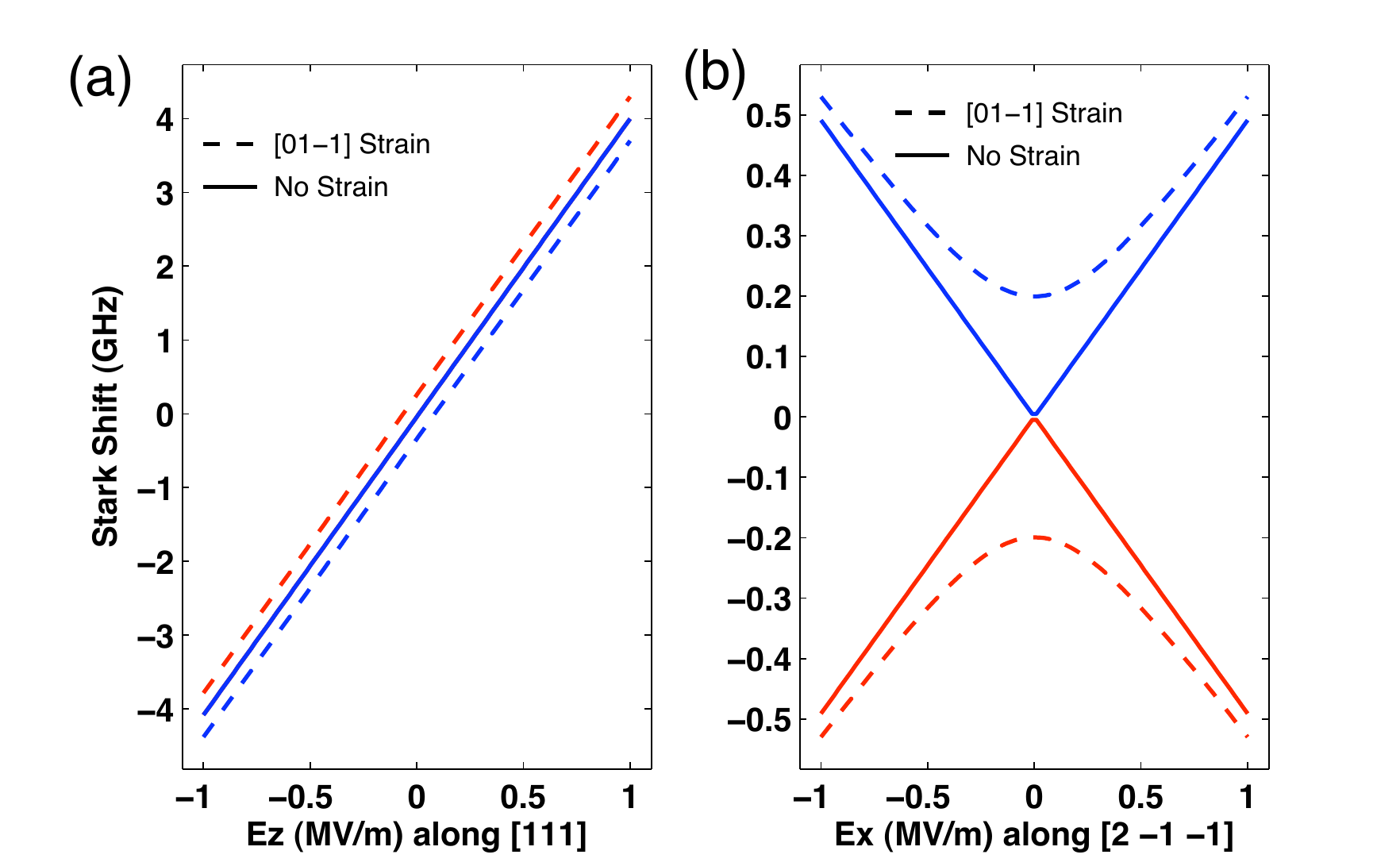}
\caption{{\bf Piezo-electric response of optical transitions.} (a)
  response to electric field $E_z$ along the NV-axis ([111] orientation or
  equivalents). The defect only shows linear Stark Shift independent
  on the initial strain. (b) electric field $E_x$ applied
  perpendicular to the NV-axis in the absence of strain (solid
  lines). The optical transitions ${^3A}_2(m_s=0)\rightarrow E_x
  (m_s=0)$ and ${^3A}_2(m_s=0)\rightarrow E_y (m_s=0)$ are split
  linearly and evenly. In the presence of strain along the $\hat{y}$
  direction (dashed lines), the response is quadratic due to the
  splitting between $E_x$ and $E_y$ states in the excited state. Our
  numerical results are in fair agreement with experimental
  results\cite{Tamarat:PRL2006}.}
\label{fig:piezo}
\end{center}
\end{figure}

The application of an electric field to a defect leads to two main
effects. The first effect, the {\it electronic effect}, consists of
the polarization of the electron cloud of the defect, and the second
one, the {\it ionic effect}, consists of the relative motion of the
ions. It has been shown that the two effects are indistinguishable, as
they have the same symmetry properties \cite{Bates:JPC1968}. The ionic
effect is related to the well-known piezoelectric effect. When a
crystal is under stress, a net polarization $P_i = d_{ijk}\sigma_{jk}$
is induced inside the crystal, where $d_{ijk}$ is the third-rank
piezoelectric tensor and $\sigma_{jk}$ represents the magnitude and
direction of the applied force. Conversely, the application of an
electric field might induce strain given by $\epsilon_{jk} =
d_{ijk}E_i$, where $E_i$ are the components of the electric field
\cite{Nye:1985}. The tensor $d_{ijk}$ transforms as the coordinates
$x_ix_jx_k$ and, therefore, group theory can be used to establish
relations between its components for a given point group. In
particular, the non-zero components should transform as the
irreducible representation $A_1$. By projecting $d_{ijk}$ (or
$x_ix_jx_k$) onto the irreducible representation $A_1$, we can
determine the non-zero free parameters of the tensor $d$ and determine
the effect of electric field on the eigenstates of the unperturbed
defect (see Appendix \ref{app:strain}). In the case of the NV center,
the effect on the excited state triplet is given by following matrix,
\begin{eqnarray} \label{eq:hee}
H_{E} = g(b+d)E_z +ga
\left( \begin{tabular}{cc|cc|cc}
$$  &  &  &  & $E_x$ & $-iE_y$ \\ 
 &  $$ & &  & $-iE_y$ & $E_x$ \\ \hline
 &  &  $E_x$ & $E_y$  & & $$ \\
 & & $E_y$ & $ -E_x$  & $$ & \\ \hline
$E_x$ & $iE_y$ & &  & $$ & \\ 
$iE_y$ & $E_x$ & $$ & & &   $$
\end{tabular}
\right),
\end{eqnarray}
in the basis $\kp{A_1, A_2, E_x, E_y, E_1, E_2}$, while the effect on the ground state triplet is 
\begin{eqnarray} \label{eq:heg}
H_{E} = 2gbE_z,
\end{eqnarray}
in the basis $\kp{{^3}A_{2+},{^3}A_{20},{^3}A_{2-}}$. The parameters
$a$, $b$ and $d$ are the components of the piezo electric tensor
$d_{ijk}$ and $g$ is the coupling between the strain tensor $e$ and
the NV center. Comparing Eq. (\ref{eq:hee}) and (\ref{eq:heg}), we
note that the linear response of the excited state and ground state
are in principle different. An electric field along the $\hat{z}$
(NV-axis) can be used to tune the optical transition without
distorting the $C_{3v}$ symmetry of the defect, provided $b\neq d$. In
Figure \ref{fig:piezo}a we show the linear response of NV centers
under an electric field parallel to the NV-axis. In this case, the
linearity is not affected by the presence of strain. Our estimates for
the ionic effect, based on the response of the lattice defect to
electric field and the response of the orbital energies to strain (see
Appendix \ref{app:strain}), indicate that the relative shift between
the ground and excited state is about 4 GHz / MV/m. This could be very
important in schemes to entangle two NV centers optically as the
wavelength of the photons emitted from each NV center need to overlap
\cite{Beugnon:Nature2006}.
%\cite{Kimble, OrSomeone}.
In addition, an electric field with components $E_{x,y}$ can be used
to completely restore the $C_{3v}$ character of the defect. In Figure
\ref{fig:piezo}b, we show the response of optical transitions under an
electric field perpendicular to the NV axis. In this case, the
response is linear if strain is absent and quadratic if strain is
non-zero. Dashed lines show the response to an electric field when
the defect experiences a 0.3 GHz strain along the [01-1] axis. Our estimations can be used to interpret  the Stark shift observations by Tamarat et al. \cite{Tamarat:PRL2006}.

\section{Conclusions}

We have used group theory to identify, analyze and predict the
properties of NV centers in diamond. This analysis can be extended to
other deep defects in solids. A careful analysis of the properties of
a defect using group theory is essential for predicting spin-photon
entanglement generation and for controlling the properties of NV
centers in the presence of perturbations such as undesired strain. We
have shown that group theoretical approaches can be applied to
determine the ordering of the singlets in the $(e^2)$ electronic
configuration and to understand the effect of spin-orbit, spin-spin
and strain interactions.

\begin{acknowledgements} 
The authors would like to thank Phil Hemmer for fruitful discussions
and acknowledge support for NSF, DARPA and Packard Foundation. JRM
thanks Fulbright-Conicyt scholarship for support. AG acknowledges the
support of Hungarian OTKA grant K-67886, the J\'anos Bolyai program
from the Hungarian Academy of Sciences and the NHDP
T\'AMOP-4.2.1/B-09/1/KMR-2010-0002 program.
\end{acknowledgements}

\appendix

\section{Dangling bond representation and character table}\label{app:cha}

In this appendix we show in detail how to find the electronic
representation for the case of the NV center. The NV center contains a
vacancy that results in broken bonds in the system. In the tight
binding picture, this means that three C atoms and one N-atom do not
have enough immediate neighbor atoms to form a covalent bond for
\emph{each} of their valence electrons. These unpaired electrons are
called 'dangling bonds'. In the case of the NV center, we consider a
simple model consisting of four $sp^3$ dangling bonds, where three of
them are centered on each of the three carbon atoms around the vacancy
and the fourth dangling bond is associated with the nitrogen atom. The
point group symmetry is $C_{3v}$ and its elements are the identity,
rotations around the $z$ (NV-axis) by $\pm2\pi/3$ and three vertical
reflection planes where each contains one of the carbons and the
nitrogen.
 
\begin{figure}[h]
\begin{center}
\includegraphics[width=0.65\textwidth]{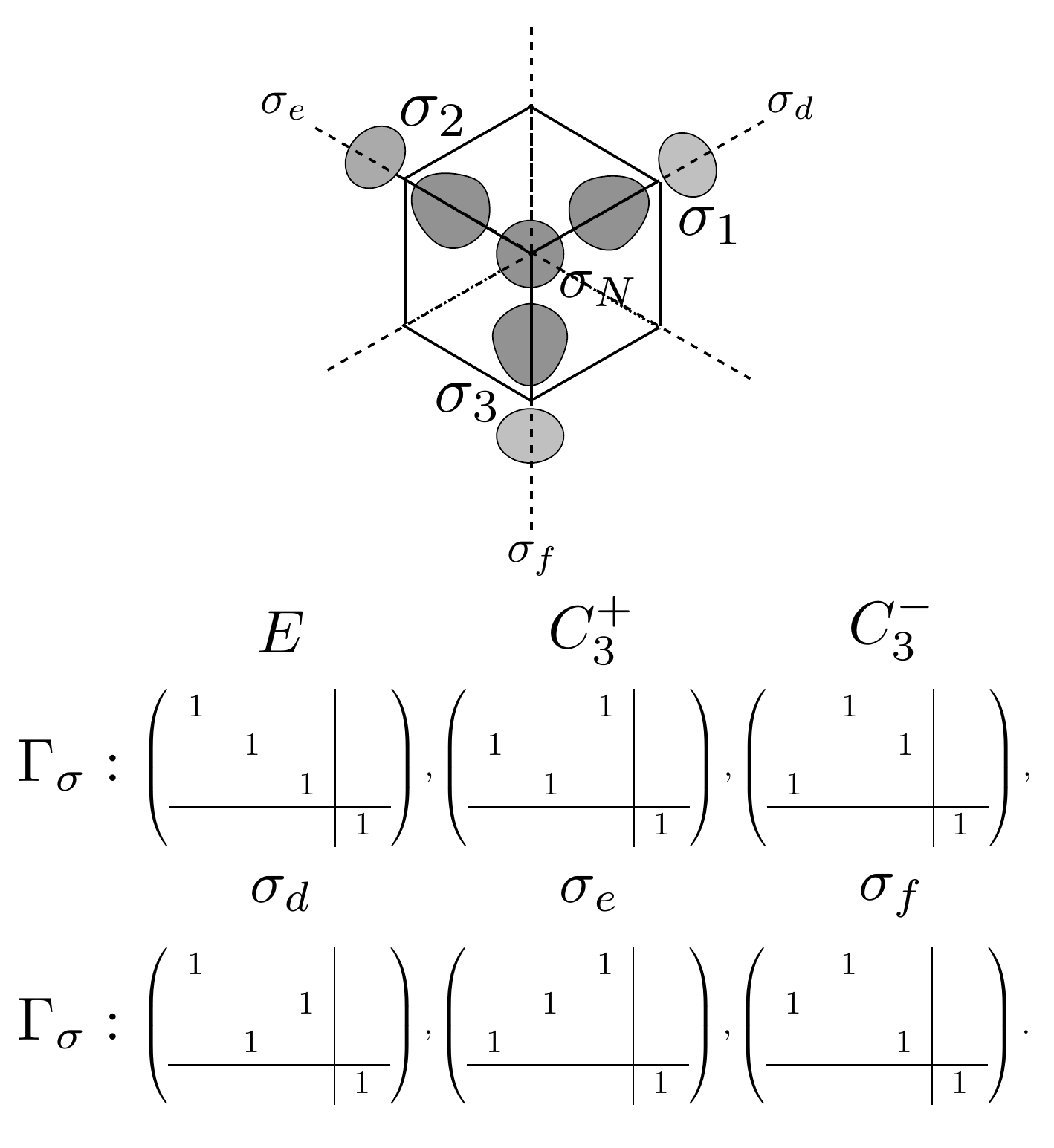}
\caption{ {\bf Schematic of the NV defect and dangling bond
    representation.} (Top) Schematics of the dangling bond orbitals
  used to represent the NV defect. The symmetry axis or NV axis is
  pointing out of the plane of the page. The dashed lines represent
  the three vertical reflections planes of the $C_{3v}$
  group. (Bottom) Matrix representation of the dangling bonds. }
\label{fig:bonds}
\end{center}
\end{figure}

As discussed in Section \ref{sec:states}, it is possible to construct
the representation of the dangling bonds for the point group they
belong to. Consider Figure \ref{fig:bonds} where the $\hat{z}$ axis is
pointing out of the paper. The dangling bonds
$\kp{\sigma_1,\sigma_2,\sigma_3,\sigma_N}$ transform into one another
under the operations of the $C_{3v}$ group. In this representation,
each operation can be written as a 4 by 4 matrix, as shown in Figure
\ref{fig:bonds}. As representations depend on the particular choice of
basis, it is customary to designate them using the trace of each matrix
(characters). Note that the character for matrices belonging to the
same class is the same, so in short the character representation for
the dangling bonds is $\Gamma_\sigma = \kp{4 1 2}$. This
representation is clearly reducible, as it can be decomposed by the
irreducible representation of the $C_{3v}$ group given in Table
\ref{t:doublec3v} \cite{Altmann:1986}.

\begin{table}[h!]
\caption{Character and bases table for the double $C_{3v}$
  group. Examples of functions that transform under a particular
  representation are $\kp{z,x^2+y^2,z^2}$, which transform as the IR
  $A_1$, the rotation operator $R_z$ as $A_2$, and the pair of functions
  $\kp{(x,y),(R_x,R_y),(xy,x^2-y^2),(yz,xz)}$ as $E$. The spin projections $\{\alpha(\uparrow),\beta(\downarrow)\}$ transform as the IR $E_{1/2}$ (or $D_{1/2}$), while the functions $\alpha\alpha\alpha+i\beta\beta\beta$ and $\alpha\alpha\alpha -i\beta\beta\beta$ transform as the IRs $^1E_{3/2}$ and $^2E_{3/2}$, respectively. }
\begin{center}
\begin{tabular}{ccccccccc}
  \hline
  \hline
  $C_{3v}$ & $E$ & $C_3$ & $3\sigma_v$ & $\bar{E}$ & $2\bar{C}_3$ & $3\bar{\sigma}_v$ \\
  \hline
  $A_1$       & 1 & 1 & 1    & 1 & 1 & 1 \\
  $A_2$       & 1 & 1 & -1   & 1 & 1 & 1  \\
  $E$         & 2 & -1 & 0   & 2 &-1 & 0  \\
  \hline
  $E_{1/2}$   & 2 & 1 & 0    &-2 &-1 & 0   \\
  $^1E_{3/2}$ & 1 & -1 & $i$ &-1 & 1 & $-i$  \\
  $^2E_{3/2}$ & 1 & -1 & $-i$&-1 & 1 & $i$   \\
  \hline
\end{tabular}
\end{center}
\label{t:doublec3v}
\end{table}%

Application of Eq. (\ref{eq:projection1}) gives the following
combination of $\sigma$'s: $\{a_C=\rp{\sigma_1+\sigma_2+\sigma_3}/3$,
$e_x=\rp{2\sigma_1-\sigma_2-\sigma_3}/\sqrt{6}$,
$e_y=\rp{\sigma_2-\sigma_3}\sqrt{2}$, $a_N = \sigma_N\}$, where $a_C$
and $a_N$ transform as the totally symmetric irreducible
representation $A_1$, and $e_x$ and $e_y$ transform as functions of
the IR $E$. Note that the $e$ states transform as vectors in the plane
perpendicular to the NV axis.

Next, we model the electron-ion interaction to find out the ordering
of these states. This interaction can be written in the basis of the
dangling bonds $\sigma_i$ as,
\begin{eqnarray}\label{eq:coulombsum}
V = v_n\ketbra{\sigma_N}{\sigma_N} + \sum_i
v_i\ketbra{\sigma_i}{\sigma_i} + h_n\ketbra{\sigma_i}{\sigma_N} +
\sum_{i>j} \ketbra{\sigma_i}{\sigma_j}h_c
\end{eqnarray} 
where $v_i<0$ is the Coulomb interaction of orbital $\sigma_i$ at site
$i$, $h_c$ is the expectation value of the interaction between
orbitals $\sigma_i$ and $\sigma_{i+1}$ at site $i=\kp{1,2,3}$, $v_n = \langle \sigma_N |V|\sigma_N \rangle$ and $h_n = \langle \sigma_i |V|\sigma_N \rangle$. This
interaction, which transforms as the totally symmetric IR $A_1$, not only sets
the order of the orbitals but also mixes orbitals $a_N$ and
$a_C$. This is a consequence of the important concept that whenever a matrix
element contains the totally symmetric representation, its expectation
value {\it might} be different from zero \cite{Tinkham:2003}. Since both wave functions as well as the interaction between them transform as the
totally symmetric representation $A_1$, the representation for the matrix element also
transform as $A_1$: $\Gamma_{\pro{}} =
\Gamma_a\otimes\Gamma_{\sigma_N}\otimes \Gamma_{int} = A_1 \supset
A_1$.  This
interaction leads to the new basis\cite{Gali:PRB2008} $\{a_1(1)=\alpha
a_c+\beta a_n$, $a_1(2)=\alpha a_n+ \beta a_c$,
$e_x=\rp{2\sigma_1-\sigma_2-\sigma_3}/\sqrt{6}$,
$e_y=\rp{\sigma_2-\sigma_3}\sqrt{2}\}$, with energies $\{E_{a_1(1),
  a_1(2)} = \frac{1}{2}(v_c+2h_c+v_n)\pm \frac{1}{2}\Delta$,
$v_c-h_c$, $v_c-h_c\}$, respectively, where $\Delta =
\sqrt{(v_c+2h_c-v_n)^2+12h_n^2}$, $\alpha^2 = 1-\beta^2 =
3h_n^2/\Delta E_{a_1(1)}$. We see that the most symmetric state is
lowest in energy, which is usually the case for attractive
interactions.

\section{Ordering of singlet states}\label{a:ordering}

Here we show that two states belonging to the same irreducible
representation should have the same expectation value for their
Coulomb interaction. We first note that the expectation value of an
operator is a scalar and it should not depend on the
particular coordinate system in use. In particular, this expectation
value should be invariant under any operation of the $C_{3v}$ group of
the NV center. The Coulomb interaction is totally symmetric and
therefore not affected by any rotation, and the wavefunctions
$\kp{e_x, e_y}$ transform as the irreducible representation
$E$. Therefore, we can get more information about these expectation
values by projecting them on the totally symmetric irreducible
representation $A_1$,
\begin{eqnarray}
(ab,V,cd) = \frac{1}{h}\sum_{R=1}^h \chi_e\rp{P_R(a)P_R(b),V,P_R(c)P_R(d)}.
\end{eqnarray}
We find as expected that
\begin{eqnarray}
({^1E}_1, V,{^1E}_1) = \frac{1}{2}({^1E}_1, V,{^1E}_1) + \frac{1}{2}({^1E}_2, V,{^1E}_2),
\end{eqnarray}
which means that the states $({^1E}_1, V,{^1E}_1)$ and $({^1E}_2,
V,{^1E}_2)$ have the same energy, as required by symmetry.

\section{Spin-spin interaction}\label{app:spinspin}

In order to analyze the effect of spin-spin interactions (Eq.
\ref{eq:spinspin}) from the perspective of group theory, we first
rewrite this interaction to identify spatial and spin terms that
transform as IR objects in the point group,
\begin{eqnarray}
h_{ss} &=& -\frac{\mu_0g^2\beta^2}{4\pi}\left[ \frac{1 -
    3\hat{z}^2}{4r^3}(s_{1+}s_{2-} + s_{1-}s_{2+}-4s_{1z}s_{2z})
  \right. \nonumber\\ &&+\frac{3}{4}\frac{\hat{x}^2-\hat{y}^2}{r^3}
  (s_{1-}s_{2-} + s_{1+}s_{2+})
  \nonumber\\ &&+i\frac{3}{2}\frac{\hat{x}\hat{y}}{r^3} (s_{1-}s_{2-}
  - s_{1+}s_{2+})\nonumber \\ &&+\frac{3}{2}\frac{\hat{x}\hat{z}}{r^3}
  (s_{1-}s_{2z} + s_{1z}s_{2-}+ s_{1+}s_{2z} +
  s_{1z}s_{2+})\nonumber\\ &&
  \left. +i\frac{3}{2}\frac{\hat{y}\hat{z}}{r^3} (s_{1-}s_{2z} +
  s_{1z}s_{2-} - s_{1+}s_{2z} - s_{1z}s_{2+})\right]\nonumber,
\end{eqnarray}
where $\hat{x},\hat{y}$ and $\hat{z}$ are directional cosines and
$s_{\pm} =s_x \pm is_y$. In the case of $C_{3v}$, for the unperturbed
center, the expectation values of the 4th and 5th terms are nonzero
in the spatial manifold of the excited state $\{\ket{X},\ket{Y}\}$
because the center lacks inversion symmetry. However, these terms might be neglected when considering
other defects with inversion symmetry. We note now that the spatial
part of the first term transforms as the totally symmetric
representation $A_1$, while the 2nd and 3rd terms transform as the
irreducible representation $E$. The reader can check which IR these
combinations belong to by looking at the character table in the
Appendix \ref{app:cha}.  Therefore, their expectation values can be written as
\begin{eqnarray}\label{eq:spacemap}
\frac{\mu_0}{4\pi}g^2\beta^2\left\langle \frac{1 - 3\hat{z}^2}{4r^3}
\right\rangle & = & \Delta( \ketbra{X}{X} + \ketbra{Y}{Y}
)\nonumber\\ \frac{\mu_0}{4\pi}g^2\beta^2\left\langle
\frac{3\hat{x}^2-3\hat{y}^2}{4r^3} \right\rangle & = & \Delta^\prime(
\ketbra{X}{X} - \ketbra{Y}{Y}
)\\ \frac{\mu_0}{4\pi}g^2\beta^2\left\langle
\frac{3\hat{x}\hat{y}+3\hat{y}\hat{x}}{4r^3} \right\rangle & = &
\Delta^\prime( \ketbra{X}{Y} + \ketbra{Y}{X} ),\nonumber \\ \frac{\mu_0}{4\pi}g^2\beta^2\left\langle
\frac{3\hat{x}\hat{z}+3\hat{z}\hat{x}}{4r^3} \right\rangle & = &
\Delta^{\prime\prime}( \ketbra{Y}{Y} - \ketbra{X}{X} ),\nonumber \\ \frac{\mu_0}{4\pi}g^2\beta^2\left\langle
\frac{3\hat{z}\hat{y}+3\hat{y}\hat{z}}{4r^3} \right\rangle & = &
\Delta^{\prime\prime}( \ketbra{X}{Y} + \ketbra{Y}{X} ),\nonumber 
\end{eqnarray}
where $\ket{X}$ and $\ket{Y}$ are the two electron states given in
Table \ref{t:states}. Note that, for symmetry reasons, the second and
third relations are characterized by the same parameter $\Delta'$,
while the last two relations are characterized by the same parameter
$\Delta''$. Similarly, it is possible to write the spin operators in
the spin basis of the two holes, $\{\ket{\alpha\alpha},
\ket{\alpha\beta}, \ket{\beta\alpha}, \ket{\beta\beta}\}$. For example,
$s_{1+}s_{2-} = \ketbra{\alpha\beta}{\beta\alpha}$. Using these
relations and Eq. (\ref{eq:spacemap}), the Hamiltonian in the
fundamental bases of the excited state of the NV center is
\begin{eqnarray}\label{eq:spinspinxy}
H_{ss} &=& -\Delta(\ketbra{X}{X} +
\ketbra{Y}{Y}) \nonumber\\ &&\otimes\left(\ketbra{\alpha\alpha}{\alpha\alpha}+\ketbra{\beta\beta}{\beta\beta}
- 2\ketbra{\alpha\beta+\beta\alpha}{\alpha\beta+\beta\alpha}\right)
\nonumber\\ &&-\Delta'(\ketbra{X}{X}-\ketbra{Y}{Y})\otimes
(\ketbra{\alpha\alpha}{\beta\beta}+\ketbra{\beta\beta}{\alpha\alpha})\nonumber\\ &&-i\Delta'
(\ketbra{X}{Y}+\ketbra{Y}{X})\otimes(\ketbra{\beta\beta}{\alpha\alpha}-\ketbra{\alpha\alpha}{\beta\beta})\nonumber \\
&& +\Delta'' \rp{ \ketbra{Y}{Y} - \ketbra{X}{X} }\nonumber \\ &&\otimes \rp{ \ket{\alpha\beta+\beta\alpha}\bra{\alpha\alpha-\beta\beta}  + \ket{\alpha\alpha-\beta\beta}\bra{\alpha\beta+\beta\alpha} }
\nonumber \\
&& +i\Delta'' \rp{ \ketbra{Y}{Y} - \ketbra{X}{X} }\nonumber \\ &&\otimes \rp{ \ket{\alpha\beta+\beta\alpha}\bra{\alpha\alpha+\beta\beta}  - \ket{\alpha\alpha+\beta\beta}\bra{\alpha\beta+\beta\alpha} }.
\end{eqnarray}
Finally, we can write $H_{ss}$ in terms of the
eigenstates of the unperturbed defect (see Table~\ref{t:states}). This
leads to Eq. (\ref{eq:nvspinspin}).

\section{Strain and electric field}\label{app:strain}

The effect of strain on the electronic structure of the defect can be
obtained from the effect of the electron-nuclei Coulomb interaction on
the eigenstates of the defect.  In our example, the Coulomb
interaction is given by Eq. (\ref{eq:coulombsum}). However, when the
positions of the atoms are such that the symmetry of the defect is
reduced, we should allow for different expectation values of the
matrix elements: $h_{ij} = \braket{\sigma_i}{V|\sigma_j} $ and $
h_{in} = \braket{\sigma_i}{V|\sigma_N}$. We have assumed that the self
interactions, $v_c$ and $v_n$, do not change as the electrons follow
the position of the ion according to the Born Oppenheimer approximation. To relate
the matrix elements to the ionic displacements, we can assume as a
first approximation that the electron orbitals are spherical
functions, and therefore the matrix elements can be parametrized by
the distance between ions, $h_{ij}\rp{q_i,q_j} =
h_{ij}\rp{|q_i-q_j|}$, so that we can write
\begin{eqnarray}
h_{ij}\rp{|q_{ij}|} \approx h_{ij}\rp{|q_{ij}^0|} +\left. \frac{1}{
  |q_{ij}| }\frac{\partial h_{ij}}{\partial q_{ij}} \rp{ q_{i} - q_{j}
}\right|_0\cdot \rp{\delta q_{i} - \delta q_{j}} + ... .
\end{eqnarray} 
The change in the matrix elements is linear in the atomic
displacements. In turn, the atomic displacements are related to the
strain tensor by $\delta q_{i} = e q_{i}$, and therefore the change in
the matrix element is given by
\begin{eqnarray}
\delta h_{ij}\rp{|q_{ij}|} \approx \left. \frac{1}{ |q_{ij}|
}\frac{\partial h_{ij}}{\partial q_{ij}} \rp{ q_{i} - q_{j} }^T e \rp{
  q_{i} - q_{j}} \right|_0.
\end{eqnarray} 
Under these considerations, it is straightforward to calculate the
effect of strain on the eigenstates of the defect. For simplicity, we
write here only the effect of strain on the degenerate orbitals, $e_x$
and $e_y$,
\begin{eqnarray}
\delta V = -g\rp{\begin{array}{cc}
  e_{xx}   &     e_{xy}  \\
  e_{xy}   &     e_{yy}  
 \end{array}},
\end{eqnarray}
where $g = \frac{8q}{3} \frac{\partial h_{ij}}{\partial q_{ij} }$ and
$q$ is the nearest neighbor distance between atoms. Using the
electron wavefunction obtained from \emph{ab initio} calculations (see
Appendix \ref{app:compmethods}) we estimate that $g\approx 2$ PHz (P = peta = 10$^{15}$).

The effect of electric field on the eigenstates of the defect can be
analyzed by the inverse piezoelectric effect as described in Section
\ref{sec:strainelectric}. In this appendix we show how group theory
can tell us the nature of the piezoelectric tensor. By projecting
$d_{ijk}$ (or $x_ix_jx_k$) onto the irreducible representation $A_1$,
we can build the following relations,
\begin{eqnarray}
a = d_{111} = - d_{221} = - d_{122}\qquad d = d_{333} \\
b = d_{113} = d_{223} \qquad c  = d_{131} = d_{232}
\end{eqnarray}
and the $d$ tensor can be written in the following short notation 
(contracted matrix form) \cite{Nye:1985}
\begin{eqnarray}
d_{ijk} \rightarrow \rp{\begin{array}{cccccc}
a & -a &  &  & c &  \\
 & &   & c &   & -2a \\
b & b & d & & &
\end{array}}.
\end{eqnarray}
For a given electric field, we have a strain tensor of the form
\begin{eqnarray}
\epsilon = \matrixtt{aE_x+bE_z}{-aE_y}{cE_x}{-aE_y}{-aE_x+bE_z}{cE_y}{cE_x}{cE_y}{dE_z}.
\end{eqnarray}
To evaluate the magnitude of the piezo-electric response, we have used
first-principles calculations as described in Appendix
\ref{app:compmethods}. The values for the components of the piezo-electric tensor due to ionic effect are $a\approx  b\approx c \approx 0.3$ $\mu$(MV/m)$^{-1}$ and $d\approx 3$ $\mu$(MV/m)$^{-1}$.

\section{Information about the first principles methods applied in our study}
\label{app:compmethods}

To determine the values of the constants $a,b,c$ and $d$ introduced in Appendix \ref{app:strain}, we applied density functional theory (DFT) \cite{Kohn65} calculations
within a generalized gradient approximation PBE (Perdew-Burke-Ernzerhof) \cite{PBE}. In
the study of spin-orbit and spin-spin interactions we used a 512-atom
supercell to model the negatively charged nitrogen-vacancy defect in
diamond. Particularly, we utilized the \textsc{VASP} code
\cite{Kresse94,Kresse96} to determine the geometry of the defect which
uses the projector augmented wave method \cite{Blochl94,Kresse99} to
eliminate the core electrons, while a plane wave basis set is employed
for expanding the valence wavefunctions. We applied the
standard \textsc{VASP} projectors for the carbon and nitrogen atoms
with a plane wave cut-off of 420~eV. The geometry optimization was
stopped when the magnitude of the forces on the atoms was lower than 0.01~eV/\AA . We
calculated the geometry of both ground and excited states. We
applied the constrained DFT method to calculate the charge density
of the excited state, that is, by promoting one electron from the $a_1(2)$
orbital to the $e_x,e_y$ orbitals as explained in
Refs.~\cite{Gali:PRL2009,Gali:PRB2009}. This procedure is a relatively
good approximation as confirmed by a recent many-body perturbation
theory study \cite{Ma:PRB2010}. The obtained geometries from
\textsc{VASP} calculations were used as starting points in the
calculations of spin-orbit and spin-spin interactions.

%The electron spin-spin interaction is not sensitive to the electron
%wave functions close to the ions, thus we applied the plane wave
%\textsc{PWSCF} code \cite{QE-2009} with ultrasoft pseudo-potentials
%\cite{Vanderbilt90} to obtain the single particle wave functions. The
%spin-spin interaction is then calculated in the Fourier-space by
%following Eq. (17) in Ref.~\cite{Rayson:PRB2008}. We used the
%occupied Kohn-Sham PBE DFT states in the gap in the evaluation of the
%electron spin-spin interaction.

The spin-orbit energy was calculated by following Eq.
(\ref{eq:so}) in our manuscript. Since the spin-orbit interaction is
short-range, we applied all-electron methods beyond the
frozen-core approximation. We utilized the \textsc{CRYSTAL} code
\cite{CRYSTAL98} for this calculation using the PBE functional within
DFT. We took the geometry as obtained from the
\textsc{VASP} calculation. We applied 6-31*G Gaussian basis set for
both the carbon and nitrogen atoms. The calculated properties (like
the position of the defect levels in the gap) agreed well with those
from plane-wave calculations.  We obtained the all-electron single particle
states and the corresponding Kohn-Sham potentials on a grid and
calculated the spin-orbit energy numerically.

Finally, we also studied the piezo-electric effect. In this case an
external electric field was applied along the NV-axis and perpendicular to
it. For this investigation only a finite size model can be used, thus we
modeled the negatively charged nitrogen-vacancy defect in a molecular
cluster consisting of 70 carbon atoms and one nitrogen atom. The
defect was placed in the middle of the cluster. The surface dangling
bonds of the cluster were terminated by hydrogen atoms. In our previous
studies we showed \cite{Gali:PRB2008} that the defect wave functions
are strongly localized around the core of the defect, thus our cluster model
can describe reasonably well the situation occuring in the bulk environment. For this
investigation we again applied DFT with the PBE functional as implemented in
the \textsc{SIESTA} code \cite{Siesta}. We used the standard double-$\zeta$
polarized basis set and Troullier-Martins norm-conserving
pseudopotentials \cite{Troullier91}. This method gives
identical results with those obtained from plane wave calculations
regarding the geometry and the wave functions in supercell models
\cite{Gali:PRB2008}. We fully optimized the defective nanodiamond with
and without the applied electric field. In this case we applied a very
strict limit to the maximum magnitude of forces on the atoms, 0.005~eV/\AA . We
applied 6 different values of the external electric field along the
NV-axis and in perpendicular directions to it, where we could clearly
detect the slope of the curvature of atomic displacements versus the
applied electric field. The resulting values for the atom displacements in
the presence of 1 MV/m electric field are on the order of a few 0.1
$\mu$\AA .

\bibliographystyle{naturemag}
\bibliography{referencesgtp,dft}

\begin{thebibliography}{10}
\bibitem{note1}
\bibinfo{howpublished}{The contact term does not contribute due to the Pauli
  exclusion principle.}
\end{thebibliography}

%Calculated with   Paper_FigureSpinSpinMay2010.m

%Calculated with Paper_GT_Fig4

%\begin{figure}[h]
%\begin{center}
%\includegraphics[width=0.75\textwidth]{Gap}
%\caption{{\bf Energy gap among states in the lower branch of the triplet excited state under strain.} a) Energy between mixtures 10 and 12 goes to zero at $\delta_{E1}^a=$ 7 GHz as well as the difference between mixtures 9-10 at 19 GHz. For such small gaps any external interaction involving spin might mix the states. b) In particular, an external magnetic field of 30 Gauss can mix the two states and create the observed anticrossing levels \cite{Tamarat:NJP2008}. Also, the nuclear spin interaction can cause a similar effect \cite{Gonzalez:PRB2009}. }
%\label{fig:gap}
%\end{center}
%\end{figure}

\end{document}